\documentclass[14pt, preprint]{aastex6}

\fullcollaborationName{The Friends of AASTeX Collaboration}
\newcommand{\msun}{~\mathrm{M}_{\odot}}
\newcommand{\zsun}{~\mathrm{Z}_{\odot}}
\usepackage{color}
\def\simpropto{\lower.2ex\hbox{$\; \buildrel \propto \over \sim \;$}}
\def\ltsim{\lower.5ex\hbox{$\; \buildrel < \over \sim \;$}}
\def\gtsim{\lower.5ex\hbox{$\; \buildrel > \over \sim \;$}}

\begin{document}

\title{Unveiling the first black holes with {\it JWST}: \\ multi-wavelength spectral predictions}

\author{Priyamvada Natarajan\altaffilmark{1, 2, 5}, Fabio Pacucci\altaffilmark{2,3}, Andrea Ferrara\altaffilmark{3}, Bhaskar Agarwal\altaffilmark{1}, Angelo Ricarte\altaffilmark{1},  Erik Zackrisson\altaffilmark{4}, and Nico Cappelluti\altaffilmark{2,5}}

\altaffiltext{1}{Department of Astronomy, Yale University, PO Box 208101, New Haven, CT 06520, USA.}
\altaffiltext{2}{Department of Physics, Yale University, P.O. Box 208121, New Haven, CT 06520, USA.}
\altaffiltext{3}{Scuola Normale Superiore, Piazza dei Cavalieri, 7 56126 Pisa, Italy.}
\altaffiltext{4}{Department of Physics and Astronomy, Uppsala University, Box 515, SE-751 20 Uppsala, Sweden.}
\altaffiltext{5}{Yale Center for Astronomy and Astrophysics, P.O. Box 208121, New Haven, CT 06520, USA.}

\% 
\begin{abstract}
Growing supermassive black holes ($\sim 10^9 \, \mathrm{\msun}$) that power the luminous $z > 6$ quasars from light seeds - the remnants of the first stars - within a Gyr of the Big Bang poses a timing challenge. The formation of massive black hole seeds via direct collapse with initial masses $\sim 10^4 - 10^5\,\msun$ alleviates this problem. Viable direct collapse black hole (DCBH) formation sites, the satellite halos of star-forming galaxies, merge and acquire stars to produce a new, transient class of high redshift objects, Obese Black hole Galaxies (OBGs). The accretion luminosity outshines that of the stars in OBGs. We predict the multi-wavelength energy output of OBGs and growing Pop III remnants at $(z = 9)$ for standard and slim disk accretion as well as high and low metallicities of the associated stellar population. We derive robust selection criteria for OBGs - a pre-selection to eliminate blue sources followed by color-color cuts $([F_{090W} - F_{220W}] > 0; -0.3 < [F_{200W} - F_{444W}] < 0.3)$ and the ratio of X-ray flux to rest-frame optical flux $(F_X/F_{444W} >> 1)$. Our cuts sift out OBGs from other infra-red bright, high and low redshift contaminants. OBGs with predicted $M_{AB} < 25$ makes them unambiguously detectable by the Mid-Infra-Red Instrument (MIRI), on the upcoming James Webb Space Telescope ({\it JWST}). For parameters explored here, growing Pop III remnants with predicted $M_{AB} < 30$ will likely be undetectable by {\it JWST}.  We demonstrate that {\it JWST}  has the power to discriminate initial seeding mechanisms.
\end{abstract}


\section{Introduction} \label{sec:intro}

The discovery of a population of bright quasars at $z \gtrsim 6$ powered by accretion onto $10^{8-10}\,\msun$ black holes (BHs, e.g. \citealt{Fan_2001, Mortlock_2011, Wu_2015}) has presented a challenge to our current understanding of supermassive black hole (SMBH) formation and growth: in particular, how these SMBHs could have assembled these masses so rapidly, within the first billion years after the Big Bang. The initial seed black holes from which these high--z quasars are predicted to form involve several distinct channels (for reviews, see \citealt{MV_2012, PN_2014, Latif_2016}) that principally involve the production of either light or massive seeds.

	In one scenario, the initial black hole seeds are essentially stellar remnants of the first stars that formed and evolved in the gas-rich early universe. The first stars in the universe, referred to as Population III (Pop III hereafter), form out of H$_2$-cooled, metal-free gas when the universe was about 200 Myr old ($z\sim 25$) onward. Pop III stars may leave behind black holes with masses $M_{\bullet,III} \sim 100\msun$, that might grow via Eddington accretion to a mass of $\sim 10^9 \msun$ by $z \sim 6$ powering the bright quasars that we now detect. This growth history though requires optimistic post-seed formation conditions, such as steady accretion powered by uninterrupted gas supply, and inefficient radiation feedback from the growing seed on the ambient gas (see e.g. \citealt{Tanaka_2012, PVF_2015, Park_2016}). There is also considerable uncertainty in the initial masses of the Pop III stars: if the initial mass function of Pop III is skewed low, a dense star cluster is expected to form, in which the mass growth of a single remnant could potentially be boosted via super-Eddington accretion as it random walks in the cluster after which it could coalesce rapidly in the environment with other individual low-mass black holes to yield a seed of $\sim 10^3 - 10^4 \msun$ (\cite{devecchi_2009, TalPN_2014}).

	Several alternate scenarios envisage the production of massive black hole seeds with initial masses that lie between $10^{3-5} \msun$ \citep{Eisenstein_Loeb_1995, Omukai_2001, Oh_2002, Bromm_Loeb_2003, Koushiappas_2004, Lodato_Natarajan_2006}. The formation of direct collapse black holes (DCBHs) is one channel that is of great current interest. This is due to the fact that the physical conditions to activate this channel are naturally available in the early universe in halos where fragmentation and star formation can be curtailed. These sites where DCBH collapse can proceed are satellite halos of copiously star-forming galaxies. Massive initial star formation leads to the production of Lyman--Werner radiation (LW) radiation with energy $11.2 \le h\nu \le 13.6\ \mathrm{eV}$. Photons in this band can photo--dissociate H$_2$ via $\rm H_2 + \gamma_{LW} \rightarrow H+H$ (see e.g. \citealt{Omukai_2001}). Atomic H is an inefficient coolant that can only cool gas down to 8000 K; H$_2$ instead can cool the gas down to $\sim$ 200 K. Therefore, the presence of an external critical LW radiation field from a neighboring early star-forming galaxy can prevent a pristine halo from forming Pop III stars. This delay, in turn, can instead lead to the isothermal collapse of the gas at 8000 K to extremely high densities, causing a runaway process that leads to the formation of a DCBH with $M_{\bullet,DC} \sim 10^{4-5}\msun$ \citep{Bromm_Loeb_2003, Lodato_Natarajan_2006, Begelman_2006, Volonteri_2008, Shang_2010, Johnson_2012, Ferrara_2014}. These more massive initial seeds can grow to SMBHs rapidly even via sub--Eddington accretion, thus circumventing the timing challenges faced by the Pop III seed scenario.

	There has been much recent progress in exploring the formation of DCBH seeds, their environment, and their early growth history \citep{Volonteri_2008, PN_2014, Agarwal_2012, Agarwal_2015, Pacucci_2015, PVF_2015}. Work thus far has suggested that light and massive seeding models can in principle be distinguished with multi-wavelength data (\cite{ET_2013, Weigel_2015}).  Attempts to do so have revealed that the most massive black holes at every epoch, that is, the massive end of the black hole mass function, were likely seeded by DCBHs \citep{Volonteri_2008, Natarajan_Volonteri_2012}.

	At high redshifts, it has been shown that there are distinct phases in the assembly history of light versus massive seeds \citep{Agarwal_2013, Volonteri_2008}. Due to the initial high masses of DCBH seeds, these black holes would have masses comparable to the stellar population in their host dark matter halos during their early mergers and evolution\footnote{The case of $J1030$ at $z=6.5$, reported by \citealt{Decarli_2012} where only the SMBH is detected prominently might be such an instance of an OBG candidate.}, in sharp contrast to hosts that harbor a light seed wherein the stellar mass would exceed the central BH mass at early times. This implies a distinctly different relationship between properties of the stellar population and the central black hole for DCBH seeds  during early epochs compared to what is seen locally \citep{Tremaine_2002, Ferrarese_Merritt_2000}. Therefore, for DCBH seeds, we infer the occurrence of an early transient stage, the Obese Black hole Galaxy stage (OBG stage; \citealt{Agarwal_2013}), during which accretion onto the DCBH is expected to outshine the stellar population, resulting in distinct, detectable spectral signatures in the infra-red and X-ray bands. Direct observational detection of rapidly growing black holes at early epochs, we argue here, is critical to probe the nature of the seeds, and direct evidence for DCBHs with a sub-dominant stellar component would be particularly telling. Here, we present the first multi-wavelength calculations of the detailed spectra for this OBG stage for growing DCBH seeds and growing Pop III seeds, with particular attention focused on features that will be detectable by instruments aboard the upcoming James Webb Space Telescope ({\it JWST}) mission. The Near-infra-red Camera (NIRCam) is sensitive in the 1 - 5 micron range, while MIRI is sensitive to detecting wavelengths from 5 - 28 microns. We note, that at the same epochs, these two seed populations have significantly different accretion histories revealed via their distinct signatures in the X-ray and IR wavelengths. 

	This paper is organized as follows. In Section~2, we present a brief overview of the massive seed formation channel and the unique properties during the OBG stage; in Section~3 we present the methodology adopted to compute the multi-wavelength spectra for growing light and massive initial seeds and the results are presented in Section~4 . We conclude with a discuss of the implications and future prospects for detection of growing BH seeds in the high-redshift universe. 

\section{The formation of DCBHs and the OBG stage}

One motivation to explore the DCBH seed formation channel is that viable physical conditions for their formation appear to be amply available in the early universe as shown by \citet{Agarwal_2012}. The dark matter halo hosts of DCBHs, as per standard structure formation in cosmology, initially start off as mini-halos with a virial temperature of $2000 < T < 10^4$ K. The first episode of star formation in these haloes could be delayed by a modest level of external LW radiation field \citep{Machacek_2001, Oshea_2008}, keeping them pristine till they grow to the atomic cooling limit $T \sim 10^4$ K. At this point, if the external LW radiation field prohibits molecular hydrogen formation by dissociation of any $\rm H_2$ that maybe present into H, DCBH formation would ensue due to the suppression of H$_2$ cooling and the inhibition of fragmentation into Pop III stars \citep{Agarwal_2012, Agarwal_2014}. Such conducive sites were found to be satellite halos in the vicinity of the first generation of star forming (SF) galaxies in the early universe (\citealt{Agarwal_2013, Agarwal_2014}). At these satellite sites, gas-rich proto-galactic disks could form that do not fragment or cool but rather go dynamically unstable (Q-unstable, see \citealt{Toomre_1964}), leading to rapid runaway accretion of gas to the center and to the formation of a massive DCBH seed. The physics and evolution of this process have been calculated analytically \citep{Lodato_Natarajan_2006, Lodato_Natarajan_2007} and conducive conditions (assembly of massive proto-galactic disks) are increasingly seen to occur in state of the art high-resolution cosmological simulations of early structure formation \citep{Regan_2009, Choi_2013, Latif_2013}. It is estimated that the typical masses of these DCBH seeds would lie in the range $10^{4-5}\msun$ and the process is limited entirely by the gas reservoir (e.g. \cite{Johnson_2012, Park_2016}; for a theoretical derivation see \citealt{Ferrara_2014}).  Furthermore, radiative feedback from these assembling DCBH seeds could halt SF in these halos \citep{Aykutalp_2014}. \cite{PFVD_2015} have shown that $> 90$\% of the gas in these proto-galactic core regions is actually accreted and the rest blown away in $T_{\rm vir} \sim 10^4$ K DCBH host halos, under radiatively inefficient conditions. Therefore, once assembled, the radiative feedback from these growing DCBHs expels gas from the host halo preventing star formation in the host halo. 

\begin{figure}[ht!]
\figurenum{1}
\epsscale{0.75}
\plotone{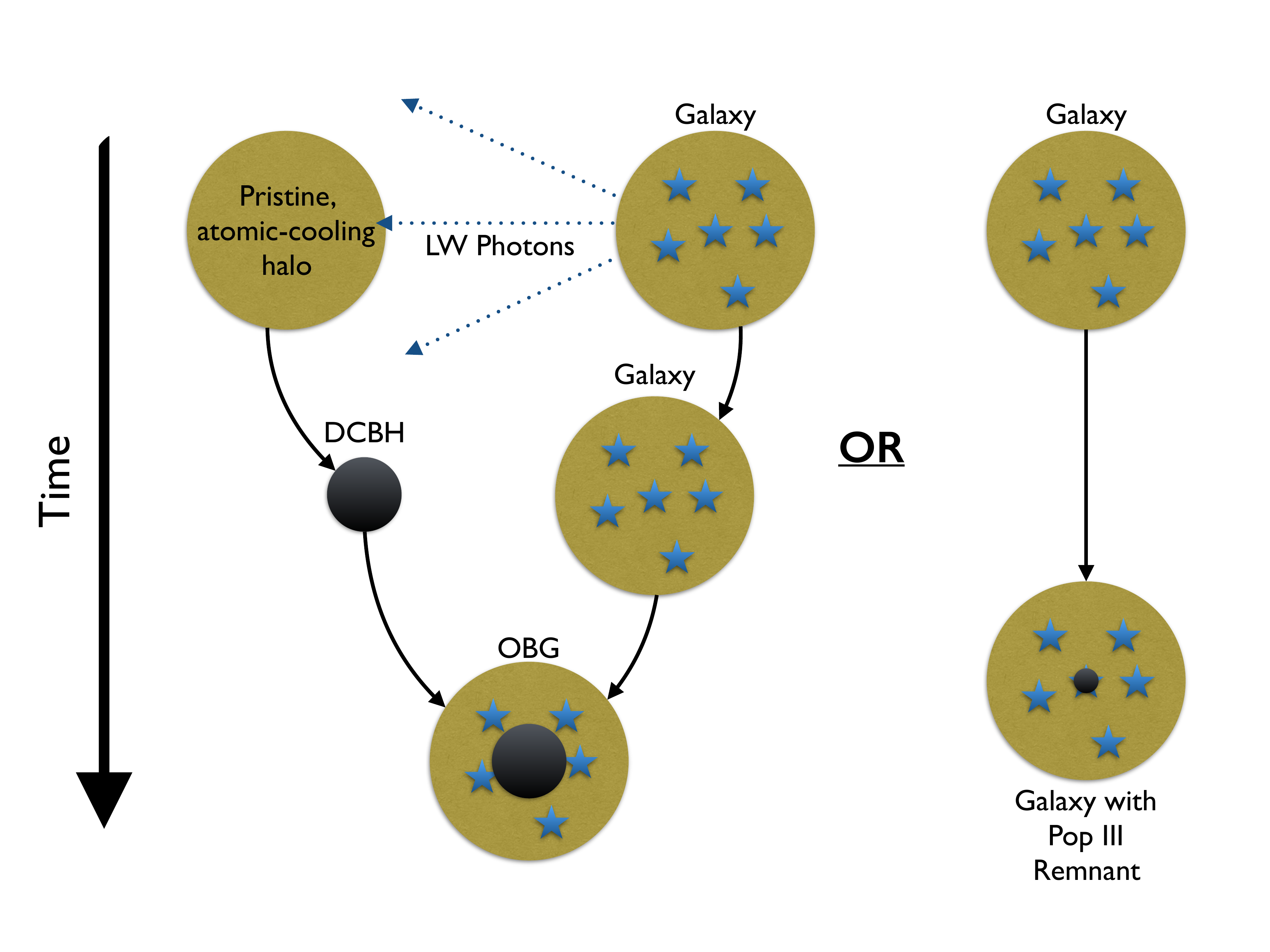}
\caption{Schematic showing the two seed scenarios and the cases studied here at $z \sim 9$ - the left hand panel: shows the formation of an OBG when a DCBH seed merges with its parent SF halo that provided the Lyman-Werner radiation that enabled its assembling. Star formation is quenched in the DCBH host halo due to the illumination by these Lyman-Werner photons, however a stellar component is accrued promptly via merging with the parent SF galaxy leading to the formation of an OBG. Here, we examine the growth following the merger of the DCBH (initial seed mass of $10^5\,\msun$) with a high and low metallicity stellar population. The right panel: shows the sequence for a growing Pop III remnant seed, which starts out with an initial mass of 100$\msun$. For these two seeding models, here we explore subsequent growth both via accretion from a standard thin disk as well as a radiatively inefficient slim disk for various metallicities of the stellar population.}
\label{fig:schematic}
\end{figure}

	Tracking merging histories of dark matter halos, we find that these DCBHs' host satellite halos rapidly merge (within $10^6\,$ yr) with the parent SF halo in their vicinity. This leads to the swift acquisition of a stellar component by the DCBH halo. This stage, corresponds to the formation of a special transitory class of objects that we have referred to as OBGs \citep{Agarwal_2013}. The OBG stage is the result of the DCBH halo merging with the LW source galaxy and acquiring its principal stellar component. This is under the assumption that star formation is halted or highly suppressed in the DCBH halo itself due to radiative feedback from the assembling seed. Meanwhile, stars and the DCBH continue to grow post merger self consistently in the sense that the same gas reservoir likely feeds the BH and forms stars in the merged remnant during this transient stage. These transitory OBGs at high redshift will lie way above the locally ($z=0$) measured correlation between the mass of the central black hole and mass of stars in the bulge where this ratio is of the order of 0.1\% (e.g. \citealt{Haring_2004}). OBGs are expected to start off well above the local $M_{BH}-\sigma$ relation and eventually proceed onto it via mergers and accretion \citep{Agarwal_2013}. This is once again distinct compared to halos that harbor lighter seeds, i.e. Pop III--black hole hosting systems, where the BH seed must grow substantially to start with and then perhaps grow in tandem with the stellar component to end up on the $M-\sigma$ relation by $z=0$. The life-cycle of Pop III remnant seeds follows a different growth pathway as the black hole accretes mass and a stellar component forms and grows in tandem all likely modulated by feedback. In this paper, we follow the growth of an OBG and a growing Pop III seed, both at high redshift ($z ~ 9$). As outlined above, the DCBH seeds form at the center of pristine, as-yet unenriched, satellite halos that are illuminated by Lyman-Werner radiation from a nearby SF galaxy (see Fig.~\ref{fig:schematic}). Star formation is quenched in these DCBH host halos, and a stellar component is accrued promptly via merging with nearby SF galaxies. In the models studied here, we examine the subsequent growth following the merger of the DCBH (initial seed mass of $10^5\,\msun$) with a neighboring high and low metallicity stellar population. The gas in the post-merger halo is assumed to have the same metallicity as the stellar population. The same scenario is explored for a growing Pop III remnant seed, which starts out with an initial mass of 100$\msun$. For both cases, we also explore growth via accretion from standard thin disk accretion as well as via a radiatively inefficient slim disk. In this treatment, we neglect the formation, merger and growth of any lower, stellar mass BHs that might form in these Pop III halos.
	
	During the OBG phase for galaxies that host a rapidly growing DCBH seed, the energy output from BH accretion vastly exceeds the emission from the stellar component. In the next section, we outline our computation of the multi-wavelength SED of an OBG under these specific circumstances. We compare and contrast the growth histories and energy output during the same epochs derived for growing Pop III seeds as well to hone in on the features in the multi-wavelength spectra that permit discrimination between these two seeding models.  

\section{Calculating multi-wavelength spectra for early black holes}

We compute the multi-wavelength SED for a growing DCBH seed in the OBG stage as well as for a growing Pop III  seed using a synthetic model to calculate the combined contribution of fluxes from the stellar and the accreting black hole components. We assemble these two components for both the DCBH and the Pop III remnant seeding models in the context of the merging hierarchy of LCDM halos using the self-consistent prescription developed in \cite{Agarwal_2012}. For instance, once a DCBH forms, its subsequent growth and energy output from accretion are simulated using a state--of--the--art 1D radiation-hydrodynamic code {\sc GEMS}. The code simulates the spherical accretion onto a high-redshift black hole seed, computing the emitted luminosity self-consistently from the accretion rate. The full spectral analysis is then performed in a post-processing step with the code {\sc Cloudy} \citep{Cloudy}. A full description of {\sc GEMS} is available in \cite{Pacucci_2015, PFVD_2015, PVF_2015}. In this work, we neglect the formation, merger and growth of any stellar mass BHs that might form in addition to the primary seed in a Pop III halo. 

	The growth and resulting energy output for the DCBH seed and lighter Pop III remnant seed are modeled in two limiting cases. In the first, that we refer to as the \textit{standard} case the growth involves accretion via a classic $\alpha$-disk \citep{Shakura_Sunyaev_1976}, which is geometrically thin and optically thick. This accretion disk is radiatively efficient and accretion occurs, on average, at the Eddington rate. As explained in \cite{PVF_2015}, this regime is entirely feedback-limited and growth gets stunted once the feedback from the accretion process heats the gas, thus disrupting the inflow. In addition to this BH growth mode, we also model the case of radiatively inefficient accretion via a \textit{slim disk} (\citealt{Paczynski_1982, Abramowicz_1988, Sadowski_2009, McKinney_2014}), in which radiation pressure is less efficient in quenching the gas inflow, due to radiation trapping. In this case, highly super-Eddington accretion rates may be reached for both massive and light seeds and the mass growth is now entirely gas supply limited. The slim disk solutions that we use capture the essential features of these complex super-Eddington flows but not all the details which require full GR-MHD treatment on the smallest scales.

The properties of the SF halo and the DCBH seed host halo that merge together are each computed using theoretical models that are built upon a suite of cosmological simulations (N--body: \citealt{Agarwal_2012}, and hydrodynamical: \citealt{Agarwal_2014}) that include physically well-motivated and self-consistent prescriptions for star formation and metal pollution from supernovae that enable the construction of detailed star formation histories (further details of the recipes used here can be found in \citealt{Agarwal_2014}). These prescriptions are used to obtain an estimate of the global and local LW radiation field needed for DCBH seeding. Here we study the accretion of metal-enriched gas onto the DCBH during the OBG stage, where the metallicity is indexed to that of the merged stellar population. At this juncture, the DCBH restarts its growth driven by the merger. In addition to the two modes of accretion outlined above, we also explore two distinct metallicities for the OBG stage: (i)   $Z = 5 \times 10^{-2}\,Z_{\odot}$ and (ii) $Z = 5 \times 10^{-4}\,Z_{\odot}$. The stellar populations that merge to form the OBG are also aged accordingly, with the lower metallicity of $5 \times 10^{-4} \zsun$ and an age of 350 Myr attributed to a younger Pop II, and a higher metallicity of $5 \times 10^{-2} \zsun$ and age of 800 Myr to an older Pop II stellar population. To model the stellar SED component of the OBG stage, we used {\sc{Yggdrasil}}, a stellar population synthesis code that employs \cite{Raiter_2010} for the lower metallicity, and {\sc{Starburst99}} \citep{Leitherer_1999} with Geneva high mass-loss tracks for the higher metallicity case. The nebular emission component was computed using {\sc Cloudy} adopting the parameters described in \cite{Zackrisson_2011}. The SEDs in all cases have been rescaled to the \cite{Kroupa_2001} stellar initial mass function.

\begin{figure}
\figurenum{2}
\epsscale{0.75}
\plotone{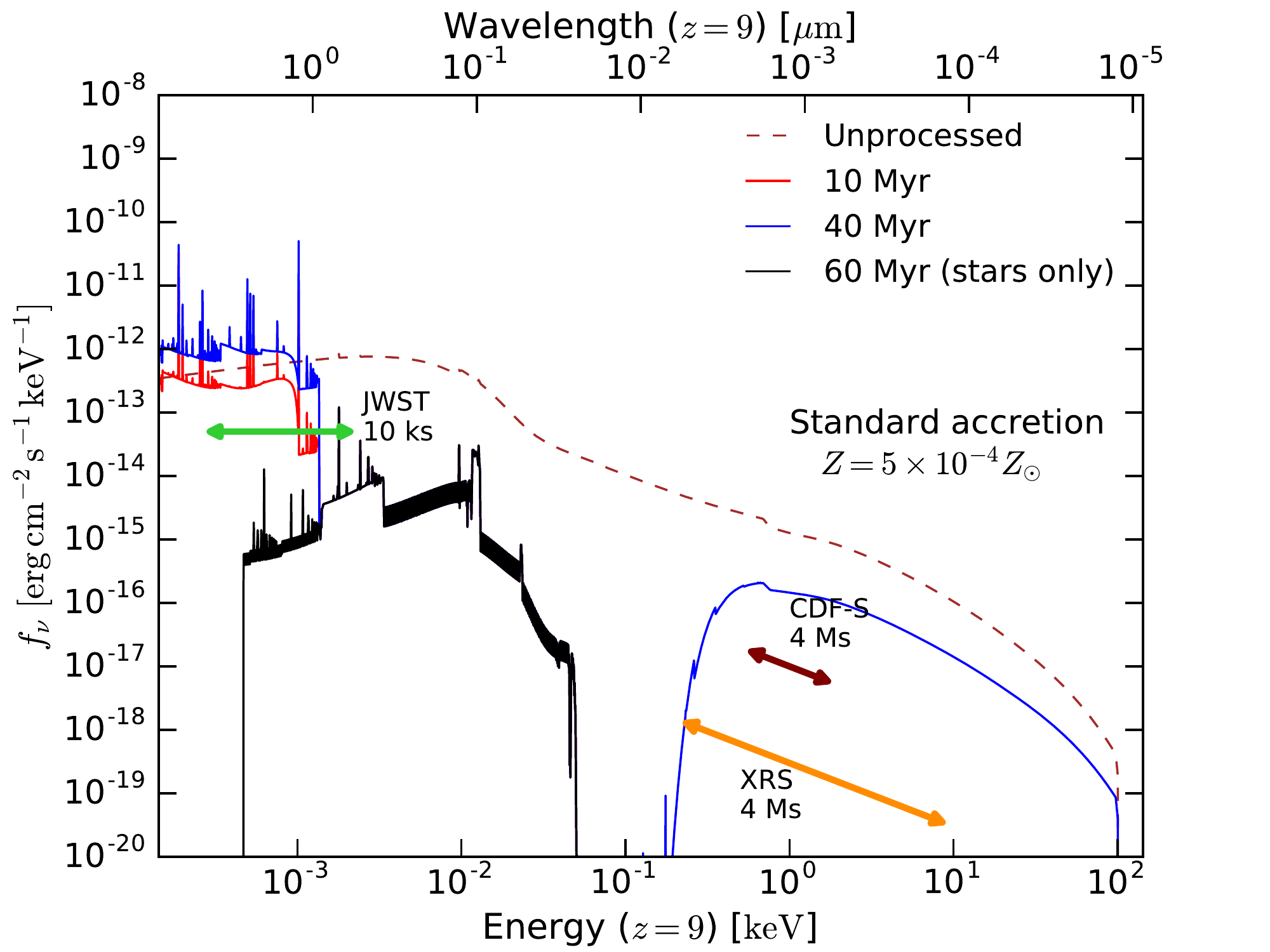}
\vspace{1.0cm}
\plotone{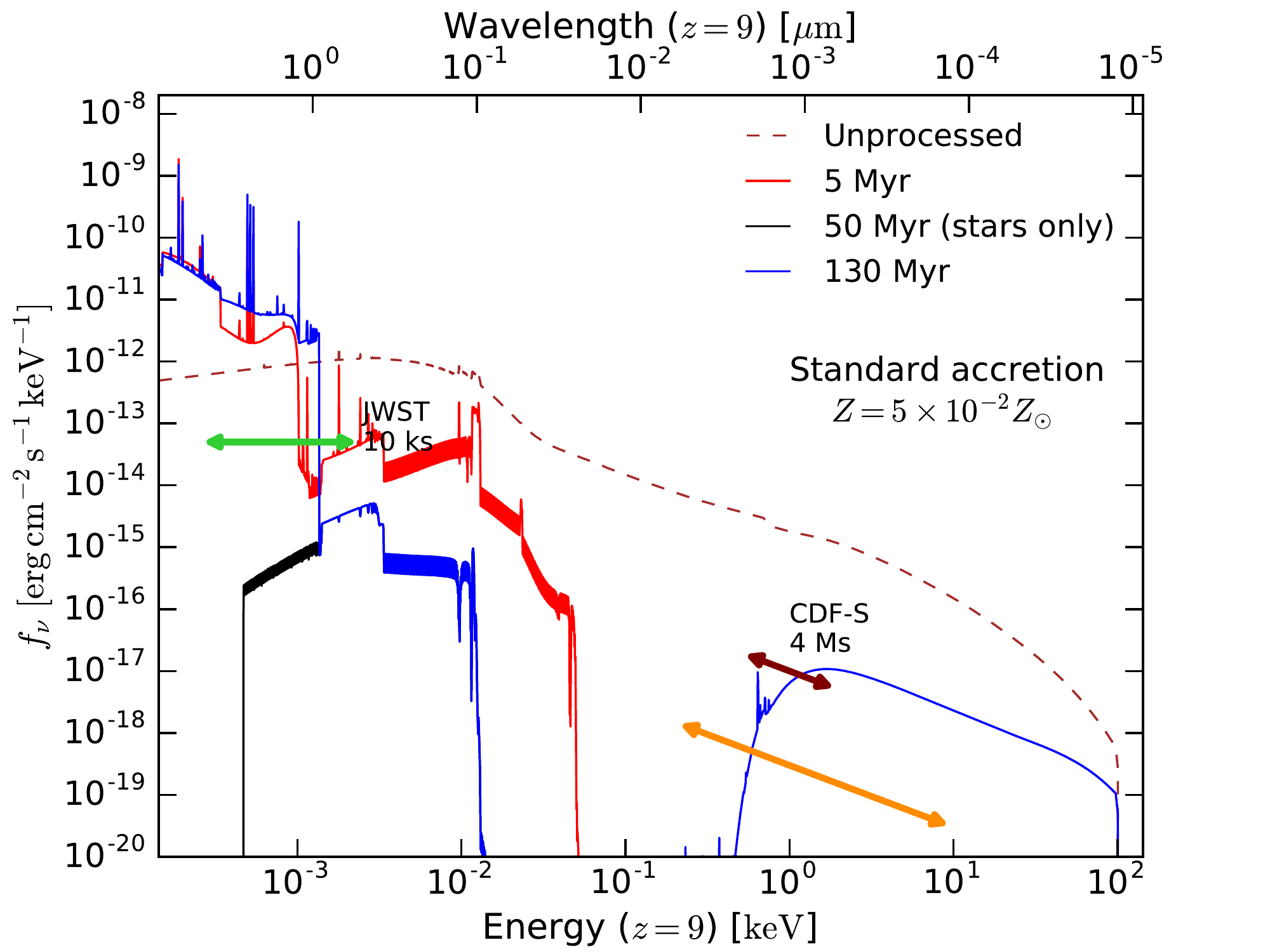}
\caption{The computed multi-wavelength OBG spectrum for growth via standard accretion for an initial $10^5\msun$ DCBH seed: In the left panel - we show the lower metallicity case ($0.0005 \zsun$) for the accreting DCBH and acquired stellar population at two time slices $t = 10\,{\rm Myr}$ and $t = 40\,{\rm Myr}$; Right panel - the higher metallicity case ($0.05 \zsun$) is plotted also at two time slices $t = 5\,{\rm Myr}$ and $t = 130\,{\rm Myr}$. The sensitivity limits for {\it JWST} for a 10 ks observation, the CDF-S (4 Ms) and the {\it X-ray Surveyor} (4 Ms) are highlighted in both panels. For contrast, we over-plot the spectrum of just the stellar component in both panels at a time slice $t = 60\,{\rm Myr}$ (high metallicity, left) and $t = 50\,{\rm Myr}$ (low metallicity, right) respectively.}
\label{fig:Std_Z2_Z3}
\end{figure}

\begin{figure}
\figurenum{3}
\epsscale{0.75}
\plotone{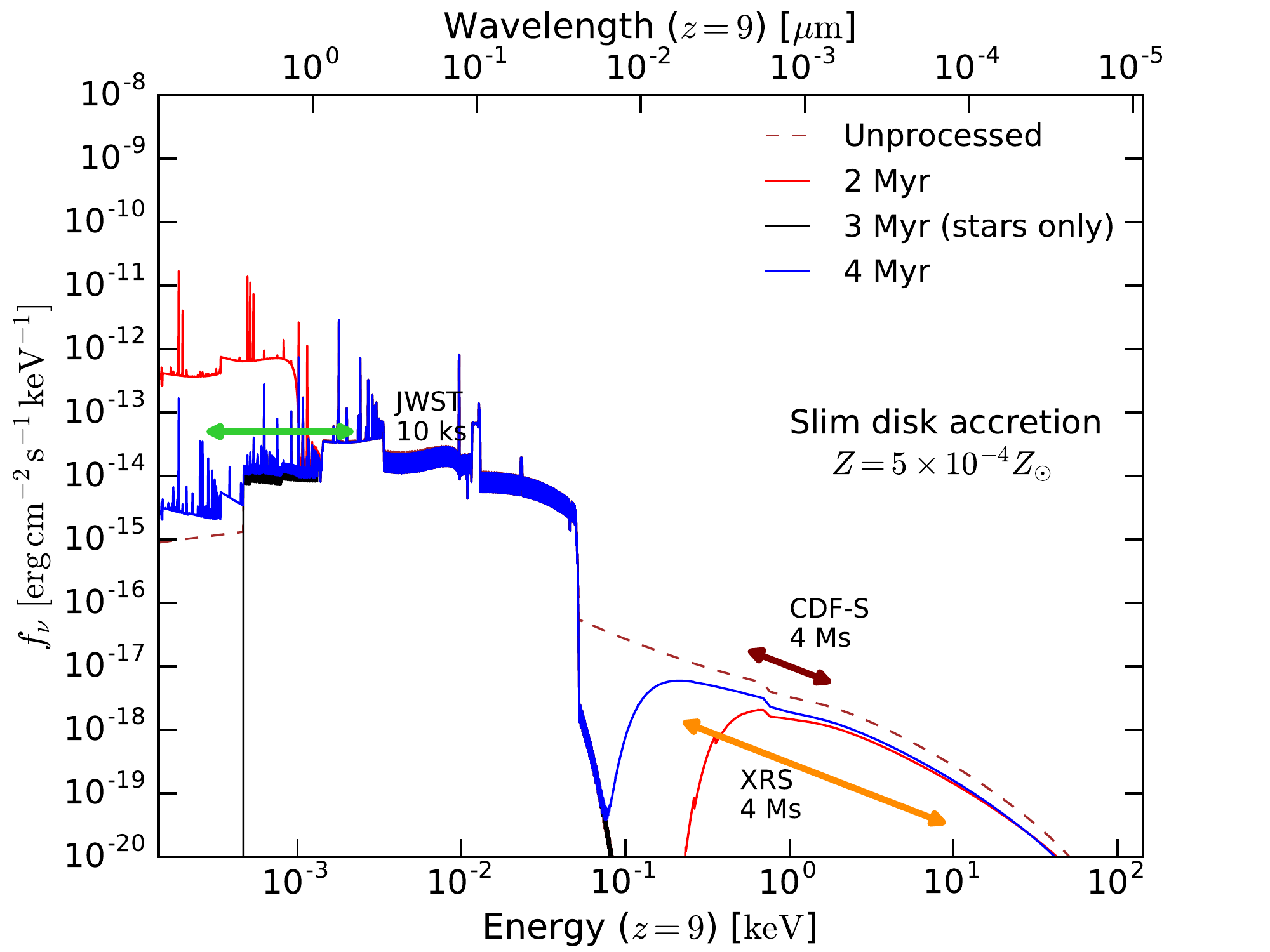}
\vspace{1.0cm}
\plotone{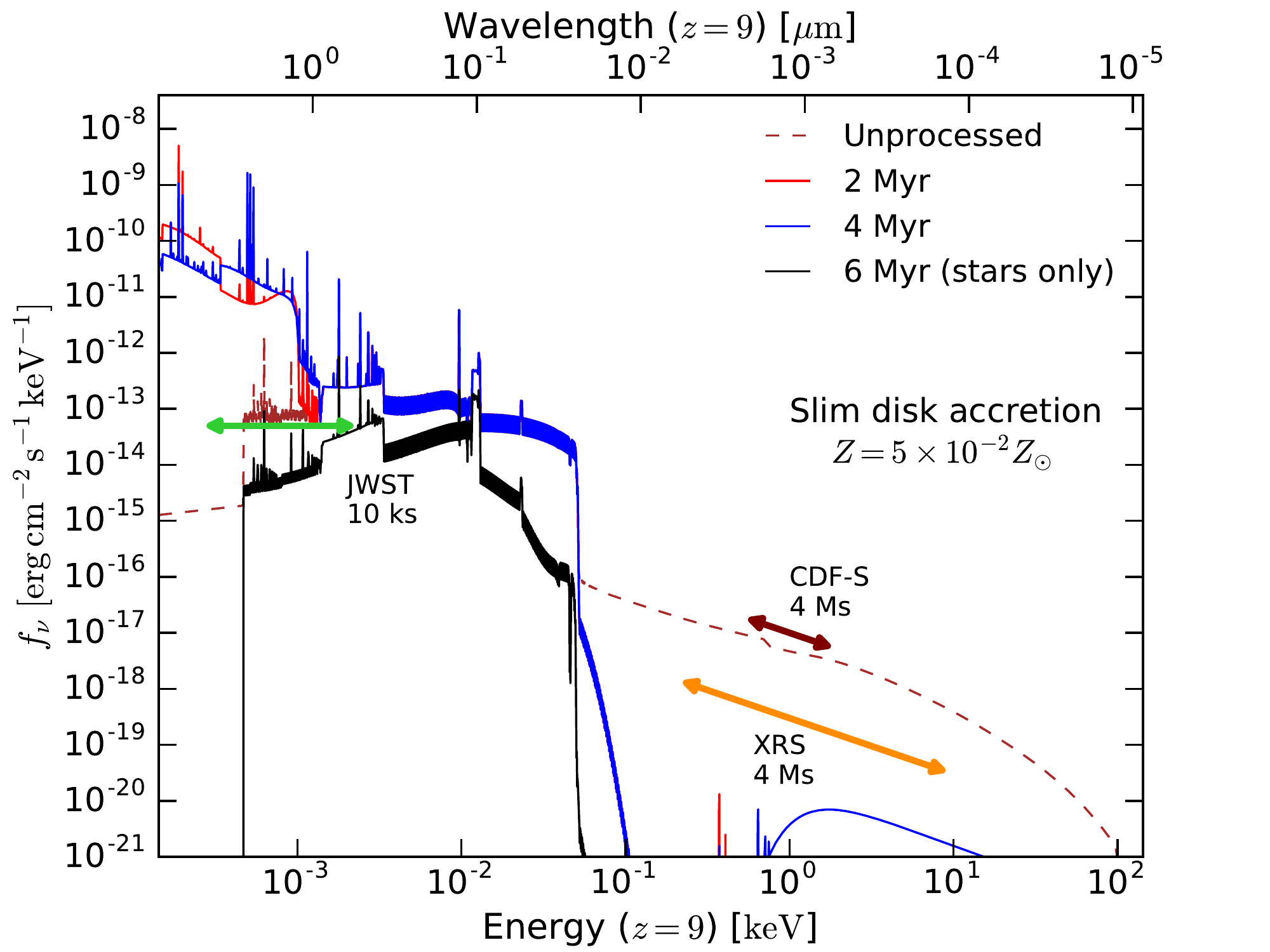}
\caption{As in Fig.~\ref{fig:Std_Z2_Z3}, with initial DCBH seed mass of $10^5\msun$, but now for growth via slim disk accretion -  in the lower metallicity case ($0.0005 \zsun$, left) at two time slices $t = 2\,{\rm Myr}$ and $t = 4\,{\rm Myr}$; and the higher metallicity case ($0.05 \zsun$, right) at two time slices $t = 2\,{\rm Myr}$ and $t = 4\,{\rm Myr}$. Once again for contrast, we over-plot the spectrum of just the stellar component at a time slice of $t = 3\,{\rm Myr}$ (high metallicity, left) and $t = 6\,{\rm Myr}$ (low metallicity, right). The reference sensitivity limits for for {\it JWST}, the CDF-S and the {\it X-ray Surveyor} are highlighted in both panels.}
\label{fig:SD_Z2_Z3}
\end{figure}

Once the satellite DCBH host halo merges with the larger host halo of the SF galaxy, the black hole thereafter is assumed to accrete from the same gas reservoir that fuels star formation. This ensures that we do not exceed the total baryon budget at any point in the growth history of the OBG. The nebular emission is also computed assuming the same ambient gas metallicity as that of the stars. We model the post--merger star formation in the OBG host halo with a constant star formation rate, leading to a final stellar mass of $M_*=10^7 \msun$ at the end of the OBG phase. The total duration of the OBG phase varies from $\sim 10 \, \mathrm{Myr}$ in the slim disk cases to $\sim 100 \, \mathrm{Myr}$ in standard accretion cases. Motivated by our prior theoretical work on the cosmological environments of OBGs \citep{Agarwal_2013}, we then combine the stellar and the accreting DCBH components, to obtain the self--consistent multi-wavelength template spectra for the various cases presented here in Figs.\ref{fig:Std_Z2_Z3}; \ref{fig:SD_Z2_Z3}; \ref{fig:PopIII_Std_Z2_Z3} and \ref{fig:PopIII_SD_Z2_Z3}. 

\vspace{1.0cm}

We also compute the multi-wavelength spectra as well as the template spectra in {\it JWST}  MIRI and NIRCam bands for the lighter Pop III seed black holes in similar conditions, namely, growth via both the standard and the slim disk mode, and in both high metallicity and low metallicity environments. Similar to the DCBH case, for the Pop III remnant starting with an initial seed mass of $100\,\msun$, we compute BH growth and star formation self-consistently till the stellar population grows to $10^7\,\msun$ as well. We follow the growth in tandem for the BH and the stellar component using the GEMS code and compute the output spectra as outlined above for the high and low metallicity cases as well as the standard and slim disk accretion mode. Once again, we model the stellar SED component of these growing Pop III seeds using {\sc{Yggdrasil}},  for the lower metallicity case, and {\sc{Starburst99}} with Geneva high mass-loss tracks for the higher metallicity case. The nebular emission component was also computed for this scenario once again using {\sc Cloudy} adopting the parameters described in \cite{Zackrisson_2011}. The SEDs were rescaled to the \cite{Kroupa_2001} stellar initial mass function. Here as well, once the Pop III remnant seed host halo merges with a larger host halo of a nearby SF galaxy, the black hole thereafter is assumed to accrete from the same gas reservoir that fuels star formation. This again ensures that we do not exceed the total baryon budget. The nebular emission is computed assuming the same ambient gas metallicity as that of the stars for this case as well. We model the post--merger star formation with a constant star formation rate yielding a final stellar mass of 
$M_*=10^7 \msun$ and comparable final BH mass. We do this to effectively compare the two seeding model cases. The results of the computed multi-wavelength spectra for these Pop III remnants seeds for the various parameter choices are presented in Figs.~4 \& 5. 

	As shown in Fig.~\ref{fig:JWST_bh_seds_withz9}, {\it JWST} has the power to discriminate between these two seeding models. In our analysis, we generate {\it JWST} photometry by integrating under each of the filters' sensitivity curves. These are available online for NIRCam and MIRI\footnote{http://www.stsci.edu/JWST/instruments/}. We say that a source is detectable if it would have been detected at a signal-to-noise ratio greater than 10 after $10^4$ seconds of observation. Exposure time and sensitivity calculators used are also available online \footnote{https://JWST.stsci.edu/science\-planning/performance\-\-simulation\-tools\-1/sensitivity\-overview}.

\section{Results}

\subsection{Observations of Early Black Holes}

As we show, the emergent spectrum from our model scenarios offers a clear discriminant of the properties of the accreting central source, between a growing Pop III seed and a more massive DCBH seed during its OBG stage. We first consider if and how a DCBH candidate can be pre-selected given its spectral shape, in currently available data from existing facilities. As pointed out by \cite{Pacucci_2016} viable DCBH candidates can be sifted out by combining {\it Hubble}, {\it Chandra} and {\it Spitzer} data. Combining the {\it WFC3} H-band (F160 W; 1536.9 nm) from {\it Hubble} with {\it Spitzer} IRAC1 (3.6 microns) and IRAC2 (4.5 microns) bands provides the color-color cuts needed. A color-color pre-selection using {\it Hubble} and {\it Spitzer}, with a diagnostic of $(IRAC1 - H)$ color versus $(IRAC2 - H)$ color, uniquely picks out potential DCBH candidates as they occupy a well defined region in this parameter space, due to their rather steep infra-red spectra (as shown in Fig.~4 of \citealt{Pacucci_2016}) with very negative $(IRAC1 - H)$ and $(IRAC2 - H)$ colors and faint H-magnitudes (typically $H \gtsim 28$ for DCBH sources at $z > 7$). In generating the output of the growing DCBH source spectrum, we note that  the UV photons, are reprocessed by the gas as infra-red emission primarily between 1 - 30 microns. Therefore, we expect {\it JWST} to be particularly efficient in finding these growing DCBH sources. In the next section, we predict template infra-red spectra, observable in the {\it JWST} NIRCam and MIRI bands (see Fig.~\ref{fig:JWST_bh_seds_withz9}). We compute and present the spectral signatures for the two black hole growth modes at low-metallicity and high-metallicity for both light (Pop III) and heavy (DCBH) black hole seeding models. We note that an increase in the metal content of the host halo modifies the emerging spectrum in two ways: (i) it increases the absorption of high-energy ($\gtsim 1\,\mathrm{keV})$ photons, and (ii) increases the power irradiated in the infra-red. The key discriminant that will help distinguish Pop III from DCBHs seeds are their infra-red colors as shown below. In addition, we note that the calculated overall normalization of the spectrum over the entire range - the estimated bolometric luminosities are actually a factor of 100 or so different. This obviously has a bearing on the viability of observation and informs our conclusion that Pop III remnants would be exceedingly hard to detect. We note here that the overall luminosity of the accretion models depends entirely on the mass of the central black hole. However, the mass of a black hole does not retain memory of its assembly history. If a source is observed at some time $t$, with luminosity $L(t)$ -  the mass $M(t)$ assuming some radiative efficiency could have been accreted onto a Pop III seed or a DCBH seed - and that would virtually indistinguishable. The difference between the two models, as we show in our work, is that the time needed for a Pop III seed to reach masses observable by the JWST are much longer than in the DCBH case. Growing DCBH seeds will be visible at higher redshifts. Besides, even as the Pop III seeds reach $10^7\, \msun$, which occurs on a much longer time scale than for the DCBH seeds, they cannot accrete at the Eddington rate since there is not much mass left in the halo. Therefore, they accrete at sub Eddington rates, due to the fact that most of the gas in the halo has been ejected through outflows, unlike the DCBH case, again due to the prolonged growth timescales. Moreover, by the time that the Pop III seed reaches a mass threshold observable by JWST, its host halo would be polluted by metals, due to star formation in the halo itself and from mergers with nearby halos. Therefore, the observation with the JWST of a Pop III seed in a quasi-metal-free host halo is virtually impossible. In this work, we grow both Pop III seeds and DCBH seeds to the same final mass $\sim 10^7 \msun$, which is the total gas mass in the original halo. For a given BH seed mass, any change in the accretion model and in the metallicity essentially amounts to a re-scaling of the total growth time needed to arrive at the final mass. The effects of changing metallicity and accretion model are, in fact, degenerate. The reason is that we obtain the same high energy absorption with a small column of high metallicity gas or with a large, obscuring column of lower metallicity gas. The metallicity of the halo gas therefore changes the overall  energy distribution due to photon scattering. A higher metallicity corresponds to an enhancement in the number of low-energy photons as higher energy photons with $E > 1$ keV are preferentially absorbed. There is also an enhancement in the re-radiated photons in the infra-red bands due to higher order Auger-like processes. Regardless of the metallicity of the stellar population which is also the adopted metallicity for the halo gas, the luminosity of the stellar population is always sub-dominant for the DCBH candidates.

\begin{figure}
\figurenum{4}
\epsscale{0.75}
\plotone{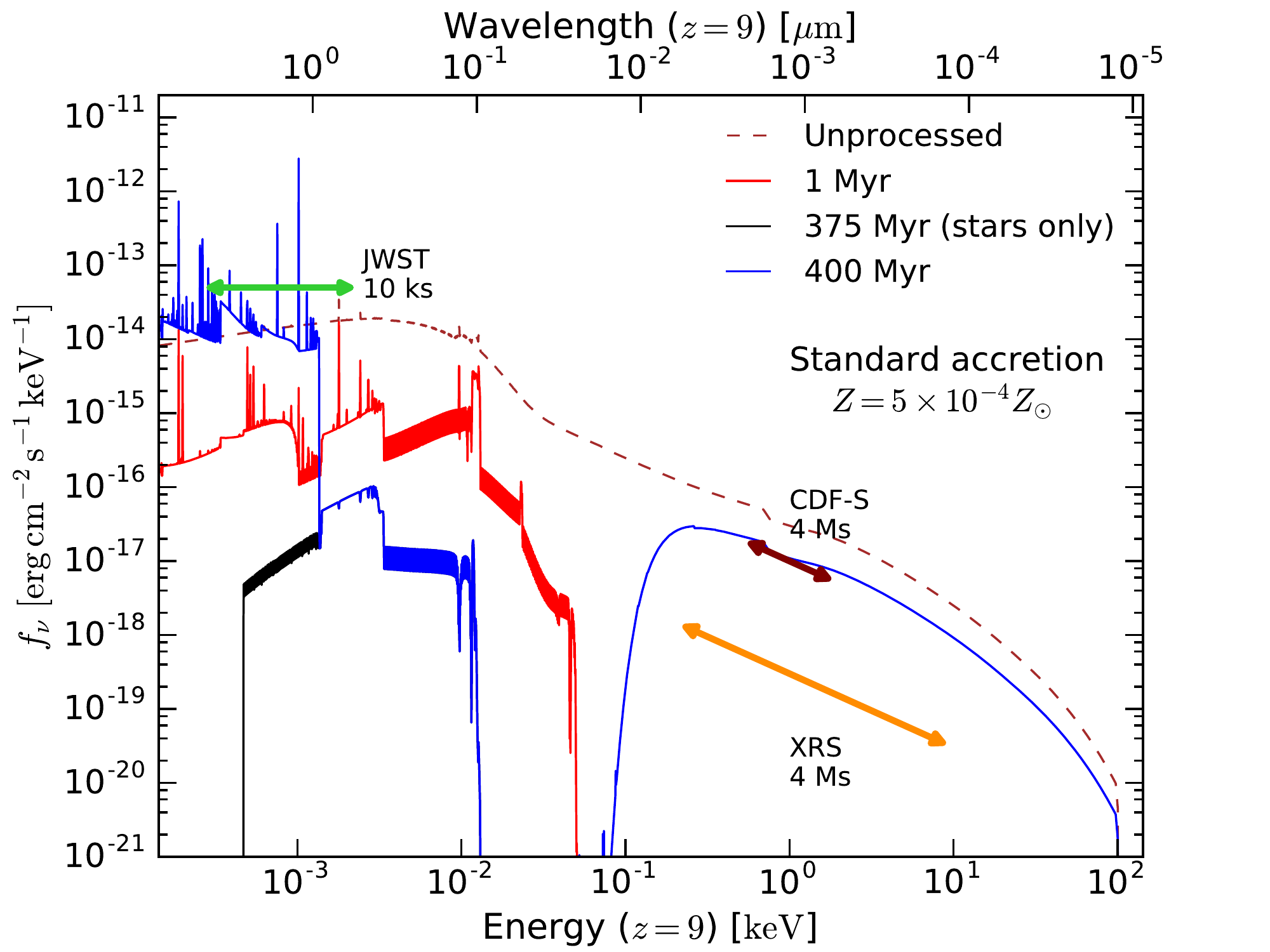}
\vspace{1.0cm}
\plotone{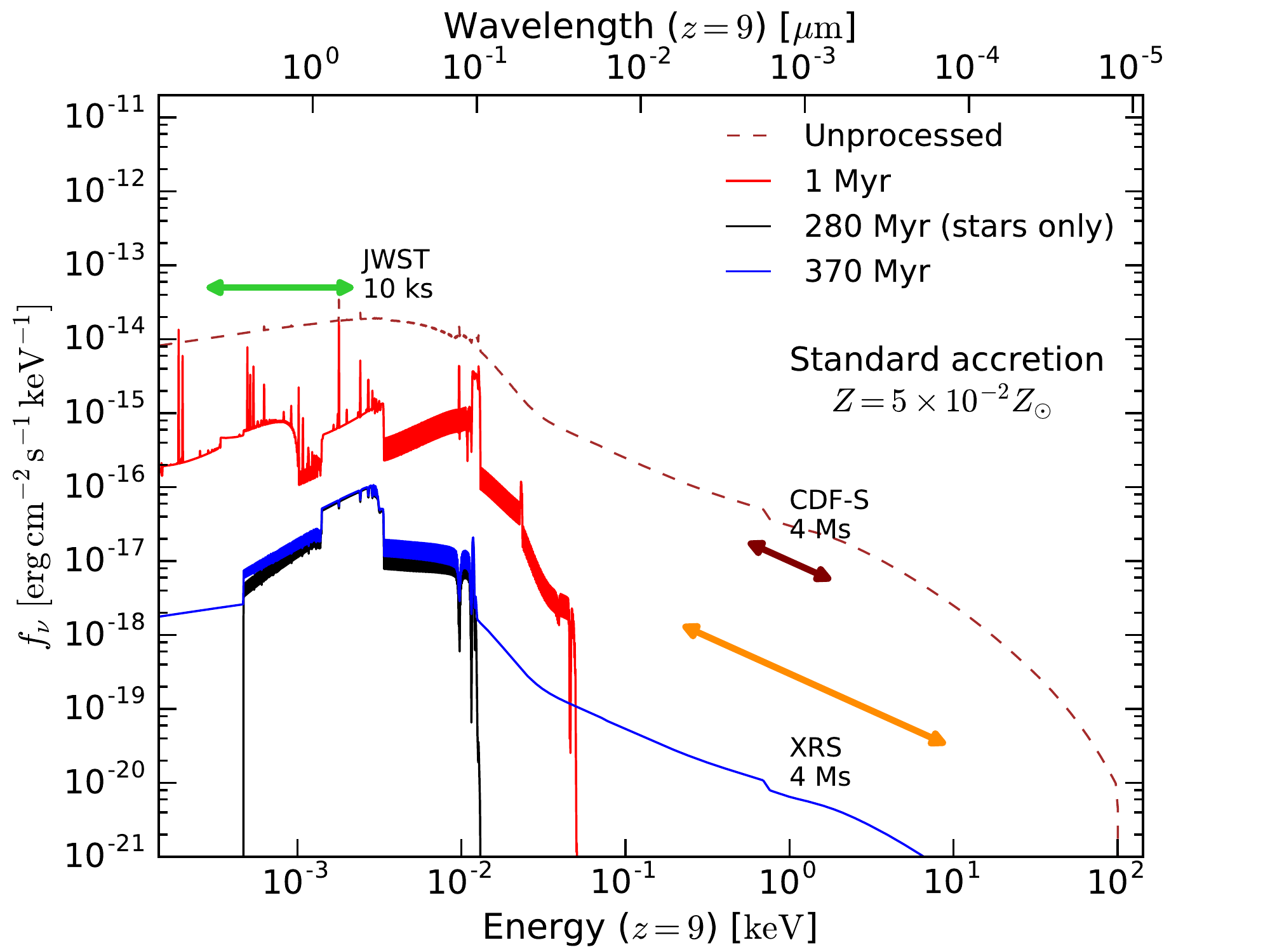}
\caption{The computed multi-wavelength spectrum for a Pop III remnant seed growing via standard accretion: in the left panel - the lower metallicity case ($0.0005 \zsun$) for the accreting Pop III and associated stellar population at two time slices $t = 1\,{\rm Myr}$  and $t = 400\,{\rm Myr}$. Over-plotted is the spectrum of the stellar population alone computed at $t = 375\,{\rm Myr}$ ; and in the right panel - the higher metallicity case ($0.05 \zsun$) also at two time slices $t = 1\,{\rm Myr}$ and $t = 370\,{\rm Myr}$. The initial mass of the Pop III seed is assumed to be $100\,\msun$. Over-plotted here in the right panel is the spectrum of the stellar population alone computed at the $t = 280\,{\rm Myr}$ snapshot. The sensitivity limits for {\it JWST}, the CDF-S and the {\it X-ray Surveyor} are highlighted in the plot. Here, we assume that only one Pop III remnant seed per halo is produced, and therefore ignore the presence of any additional stellar mass remnants.}
\label{fig:PopIII_Std_Z2_Z3}
\end{figure}

\begin{figure}
\figurenum{5}
\epsscale{0.75}
\plotone{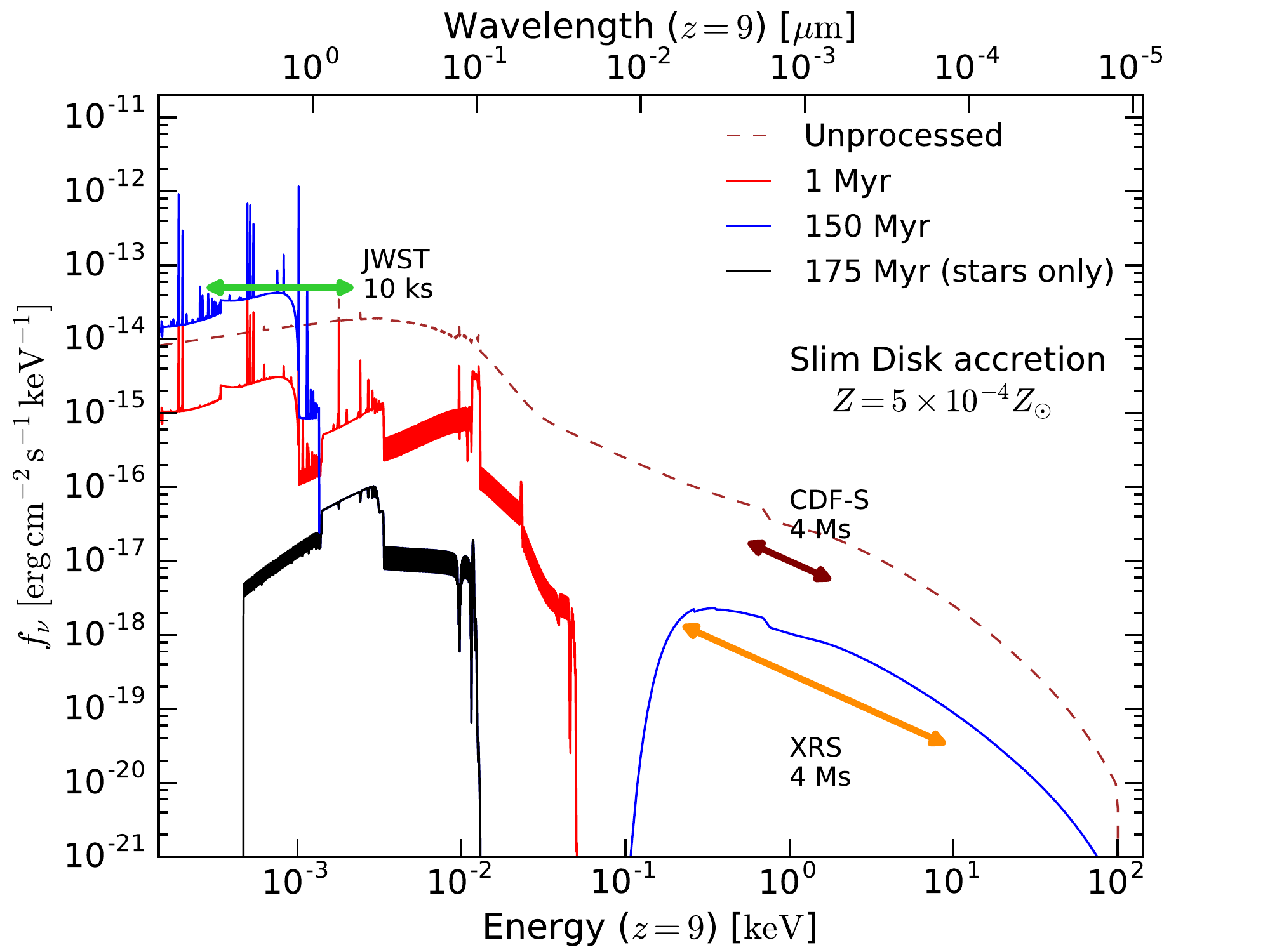}
\vspace{1.0cm}
\plotone{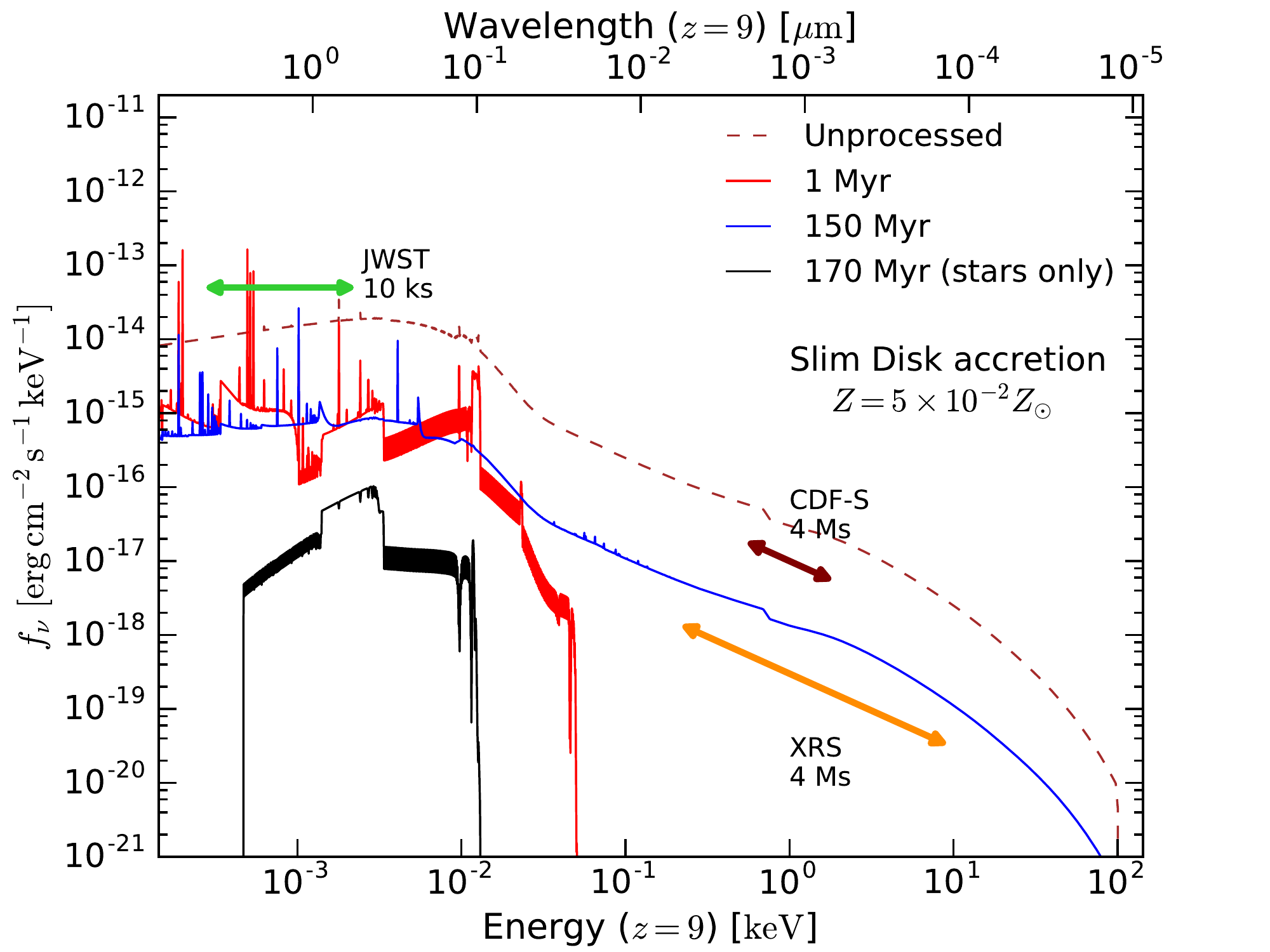}
\caption{As in Fig.~\ref{fig:PopIII_Std_Z2_Z3}, for a growing Pop III remnant seed, here we plot the computed multi-wavelength spectrum for slim disk accretion: in the left panel - computed spectrum for the lower metallicity case ($0.0005 \zsun$) at two time slices $t = 1\,{\rm Myr}$  and $t = 150\,{\rm Myr}$. Overplotted is the spectrum of just the stellar population at the $t = 175\,{\rm Myr}$ snapshot. In the right panel - the higher metallicity case ($0.05 \zsun$) at two time slices $t = 1\,{\rm Myr}$  and $t = 150\,{\rm Myr}$. In both cases, we track the growth of  a $100\,\msun$ initial mass Pop III seed, and for contrast we over-plot the spectrum of just the stellar population extracted from the model at $t = 170\,{\rm Myr}$ f. Once again, the reference sensitivity limits for for {\it JWST}, the CDF-S and the {\it X-ray Surveyor} are highlighted in both panels.}
\label{fig:PopIII_SD_Z2_Z3}
\end{figure}

Aside from the discrimination between template growing Pop III spectra and OBG candidate spectra in infra-red wavelengths, there are also clear signatures in X-ray wavelengths that we use to construct a selection methodology. From the template spectra in Fig.~\ref{fig:Std_Z2_Z3} and Fig.~\ref{fig:SD_Z2_Z3} distinct selection criteria in the X-ray as well as optical and infra-red bands can be determined to hone in on the OBG stage for DCBH seeds. We note that integrating the X-ray flux in the derived spectra over the {\it Chandra} bands at 0.5 - 2 keV and 2 - 7 keV, we find that  $\approx 10-15\%$ of the bolometric luminosity is emitted in X-rays and 20\% in the optical band for the standard accretion case and 25\% for the slim disk case. The X-ray flux for OBGs in the standard accretion case is comparable in the hard and soft bands - with flux  $\ltsim 5.0 \times 10^{-16} \, \mathrm{erg \, s^{-1} \, cm^{-2}}$ in the soft band ($0.5 - 2.0 \, \mathrm{keV}$) and $\ltsim 5.6 \times 10^{-16} \, \mathrm{erg \, s^{-1} \, cm^{-2}}$ in the hard band ($2.0 - 7.0 \, \mathrm{keV}$). As already pointed out in \cite{PFVD_2015} for a DCBH growing via the standard accretion mode, the OBG stage should already be observable by {\it Chandra}.  If however accretion occurs primarily through the highly-obscured slim disk, the expected fluxes are a factor of a few lower although they remain roughly equal in the hard and soft bands ($\ltsim 2.8 \times 10^{-16} \, \mathrm{erg \, s^{-1} \, cm^{-2}}$ in the soft band and $\ltsim 4.0 \times 10^{-18} \, \mathrm{erg \, s^{-1} \, cm^{-2}}$ in the hard band). Given these estimates, a typical DCBH growing via the slim disk mode is not detectable with current {\it Chandra} data for both the high and low metallicity cases. However, given expected design specifications for the proposed, future X-ray telescope X-Ray Surveyor, as shown in Fig.~\ref{fig:Std_Z2_Z3} and Fig.~\ref{fig:SD_Z2_Z3}, detection of these objects in the full X-ray range (hard and soft bands), independent of the metallicity of the host halo and the accretion model, is expected. Regardless of their X-ray fluxes, OBG candidates as we show are easily detected in the MIRI bands in both the high and low metallicity cases (left panel of Fig.~\ref{fig:JWST_bh_seds_withz9}). Such candidates would be characterized by a red slope between 1 and 10 microns and a flat slope beyond. This is in contrast to interloping low redshift sources that are extremely blue in the far IR as seen in Fig.~\ref{fig:JWST_bh_seds_withz9} \citep{Oesch_2016}.

\begin{figure}[ht!]
\figurenum{6}
\epsscale{0.75}
\plotone{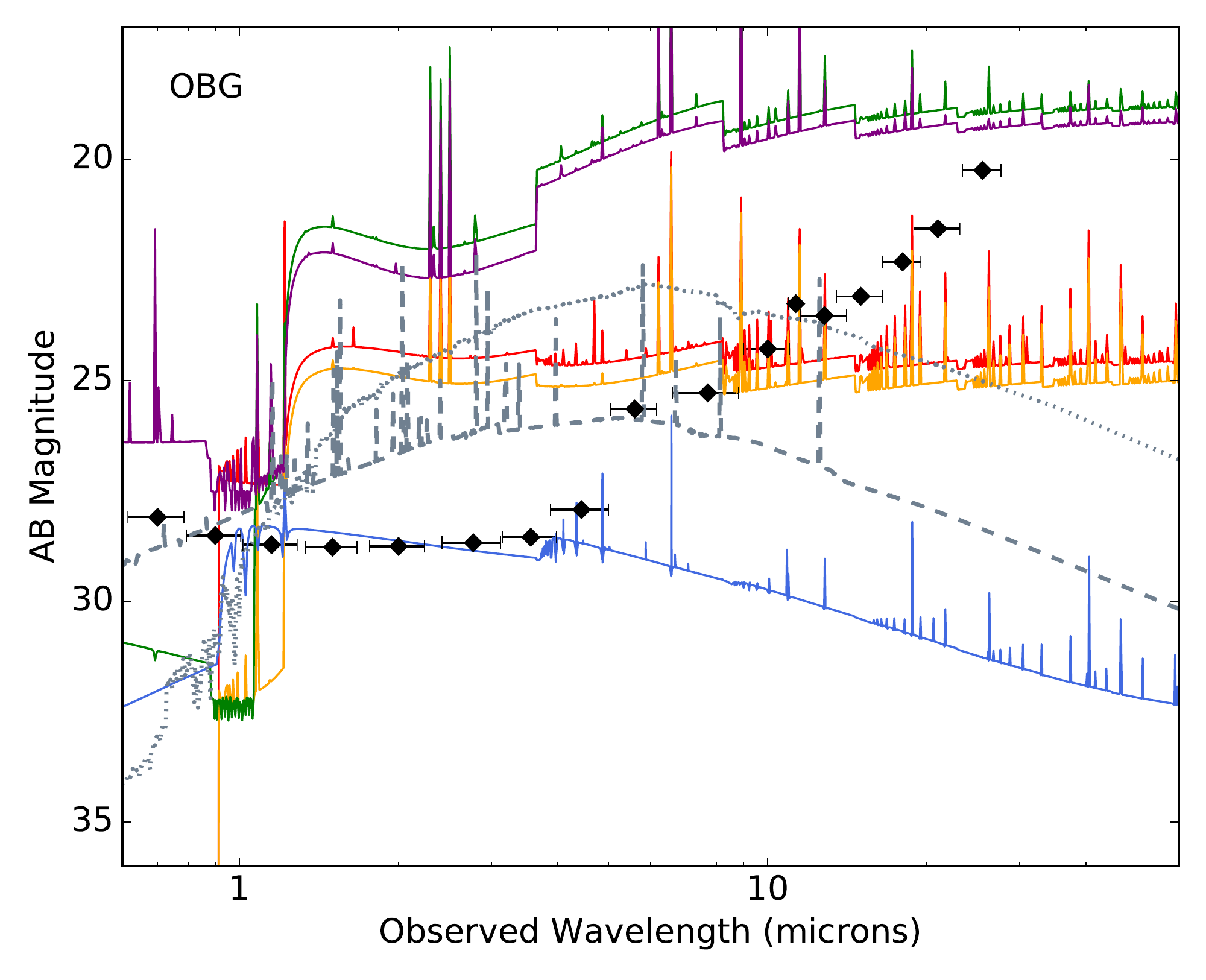}
\vspace{1.0cm}
\plotone{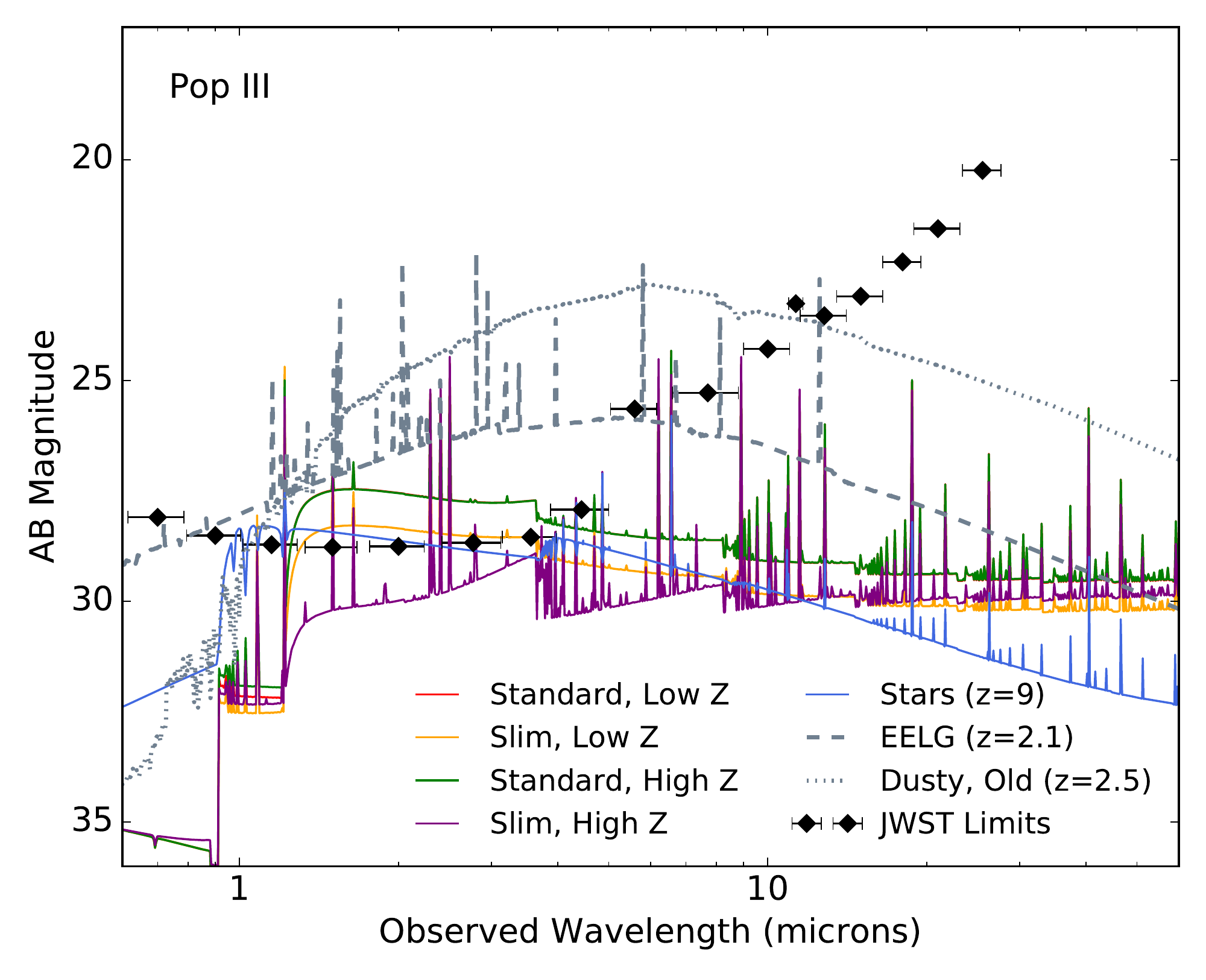}
\caption{Predicted {\it JWST} template spectra for the two black hole seeding models and other key contaminating high and low redshift sources. For both seeding models, the SED snapshots with the maximum-IR flux have been selected for plotting here. In the left panel - OBGs with an active DCBH accreting in the standard and slim disk cases for high and low metallicity at $z = 9$. Here, for standard accretion the high metallicity case is shown in green; and the low metallicity case in red. In the slim disk accretion case, the lower metallicity template is plotted in orange and the higher one in purple. In the right panel - actively accreting Pop III seeds in the standard and slim disk cases for high and low metallicity also at $z = 9$ are shown. For standard accretion the high metallicity case is shown in green; and the low metallicity case in red. In the slim disk accretion case, the lower metallicity template is plotted in orange and the higher one in purple. In both panels, for comparison the spectrum of a typical actively star-forming high redshift galaxy (blue); and the measured spectra for low-redshift interlopers - an old and dusty star forming galaxy (grey dotted) at $z = 2.5$ and an Extreme Emission Line Galaxy (EELG) at  $z = 2.1$ (grey, dashed) taken from \cite{Oesch_2016} are over-plotted. The black symbols in both panels correspond to the MIRI and NIRCam filter band passes on {\it JWST} and the error bars denote the width of the filters.}
\label{fig:JWST_bh_seds_withz9}
\end{figure}

Summarizing the spectral features for our seed candidates, accretion models and metallicites, we report:

\begin{itemize}
\item{For a growing DCBH seed with initial mass $M_{\rm seed} \sim 10^5\,\msun$:
\\
\textit{Standard (Eddington--limited) growth (Fig.~\ref{fig:Std_Z2_Z3}):} We find that the signature of the OBG stage is particularly distinct for the high metallicity case, as the overall infra-red flux is elevated compared to the lower metallicity case by about two orders of magnitude. Notably, the X-ray flux is significantly higher for the lower metallicity case. Despite their lower X-ray fluxes, the higher metallicity OBGs will be clearly detected in the NIRCam and MIRI bands. \\
\textit{Slim--disk (super-Eddington) growth (Fig.~\ref{fig:SD_Z2_Z3}):} In this relatively short lived (5-10 Myr) scenario, OBGs would have diminished X-ray emission, and enhanced flux in the infra-red for the higher metallicity case. Due to the overall lowered X-ray flux, detection will require deeper X-ray exposures than the ones currently available in the {\it Chandra} archive. Once again, despite their lower X-ray fluxes, the higher metallicity OBGs accreting via slim disks will also be detected unambiguously in the MIRI bands.} 
\\
\item{For a growing Pop III seed with initial mass $M_{\rm seed} \sim 10^2\,\msun$:
\\
\textit{Standard (Eddington--limited) growth (Fig.~\ref{fig:PopIII_Std_Z2_Z3}):} For a Pop III seed, although the emergent X-ray flux would be comparable to that during the OBG phase of a growing more massive DCBH seed, we find that the reprocessed flux output in the infra-red is suppressed by about 3-4 orders of magnitude due the overall lower bolometric luminosity. Therefore, the growing Pop III source would be undetectable by {\it JWST} during most of its evolution, even though the duration of accretion is longer than for the DCBH case, reaching a maximum of $\sim 400 \, \mathrm{Myr}$.} Compared to the DCBH seed case, the emission from the stellar component for growing Pop III is not negligible. \\
\textit{Slim--disk (super-Eddington) growth (Fig.~\ref{fig:PopIII_SD_Z2_Z3}):} In this case, both the X-ray and the infra-red fluxes are much fainter for the lighter growing Pop III seed when compared to the case of accretion onto a DCBH. Hence, the highly-obscured accretion mode onto a Pop III seed would require next-generation telescopes to be detectable.

\end{itemize}

As can be seen in the figure, a fiducial OBG spectrum can be distinguished from other known IR bright sources at high redshifts, or extremely dusty lower redshift sources, particularly in the 2 - 10 micron range. In particular, we note that the SED of an OBG differs markedly from that of other typical high redshift sources: accreting Pop III seeds (see Fig.~\ref{fig:PopIII_Std_Z2_Z3} and Fig.~\ref{fig:PopIII_SD_Z2_Z3}), and infra-red bright copiously star-forming galaxies, as well as other lower redshift galaxies with extremely old, red stellar populations.

\subsection{Searching for {\it JWST} OBG candidates with current Chandra, Spitzer and Hubble data}

We explore data from current multi-wavelength surveys to derive selection criteria that we can use to data-mine for DCBH candidates in the OBG stage as well as for potential contaminants. These were developed for sifting viable candidates from a landscape of template spectra in the viable OBG parameter space. In the previous section, we saw that DCBHs growing via slim disk accretion due to their suppressed X-ray fluxes (as can be seen in Fig.~\ref{fig:SD_Z2_Z3}, and Pop III sources no matter how they accrete, are simply not luminous enough and therefore cannot be found in existing X-ray data even in the deepest exposures currently in hand. Once again looking at X-ray wavelengths in Fig.~\ref{fig:Std_Z2_Z3}, we note that DCBHs growing via standard disks in low metallicity halos in the OBG stage are the most feasibly detected candidates and could be lurking in available deep Chandra X-ray data. We reiterate that these sources are characterized by high bolometric luminosities of the order of $\sim 10^{44}\,{\rm ergs^{-1}}$, and by mild absorption in the IR. We therefore devise three-tiered criteria involving the demarcation of a color-color cut that takes into account the Lyman break (for redshift selection) and the shape of the SED in the IR, and the ratio of X-ray to rest-frame optical flux (which corresponds to observed wavelength of 2-4.5 microns) to isolate these candidates in {\it JWST} observations. The X-ray to rest-frame optical flux ratio much greater than 1 selection picks out the presence of an actively accreting source and as we show below also helps the removal of confounding star-forming galaxies with no central AGN. As shown earlier, the color-color cut also ensures the removal of contaminating passive, dusty, star forming galaxies at low redshift. Of course, we pre-select all candidates with no detection in the blue-bands (U and B-bands).

\begin{figure}
\figurenum{7}
\epsscale{0.75}
\plotone{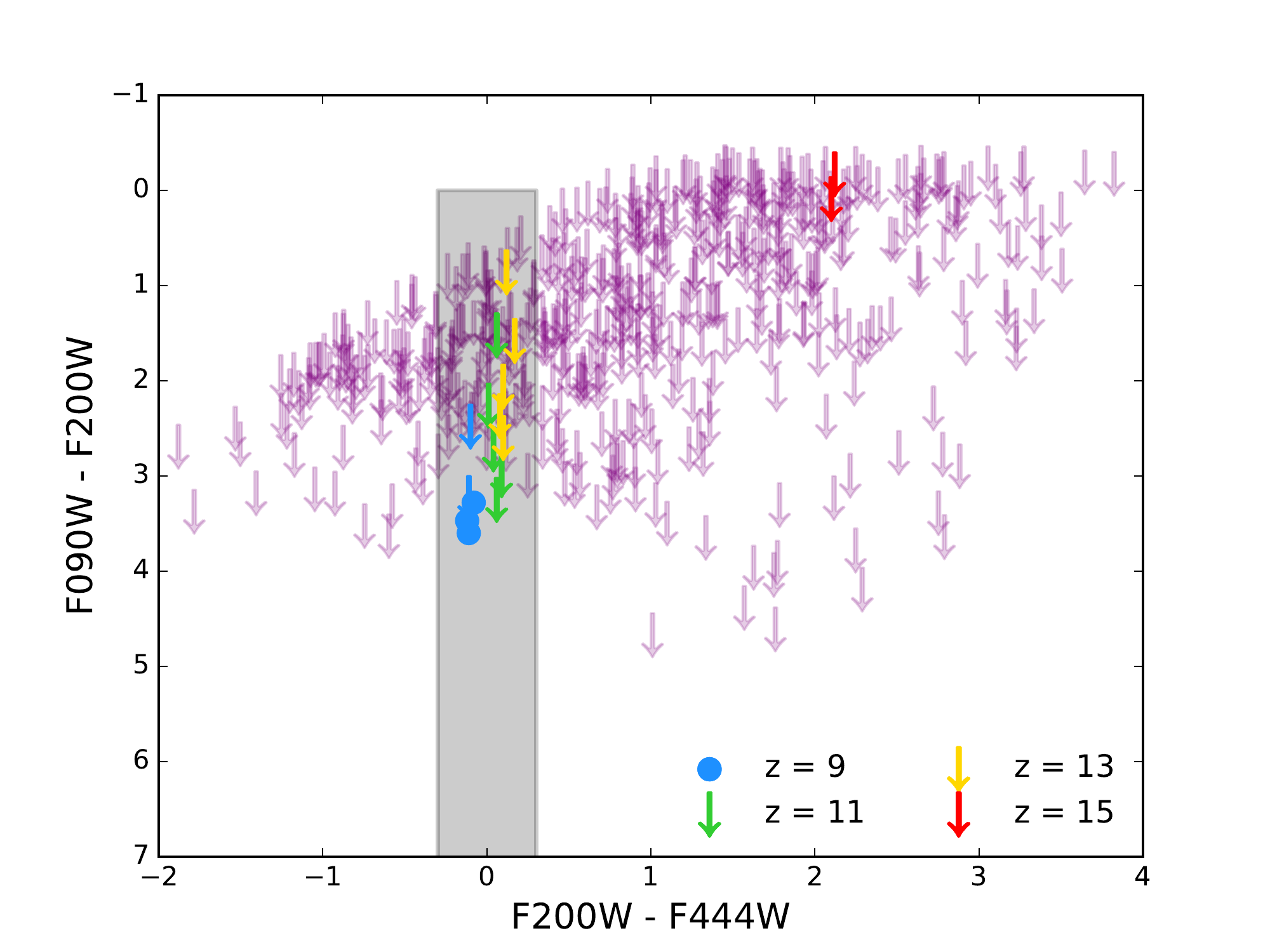}
\vspace{1.0cm}
\plotone{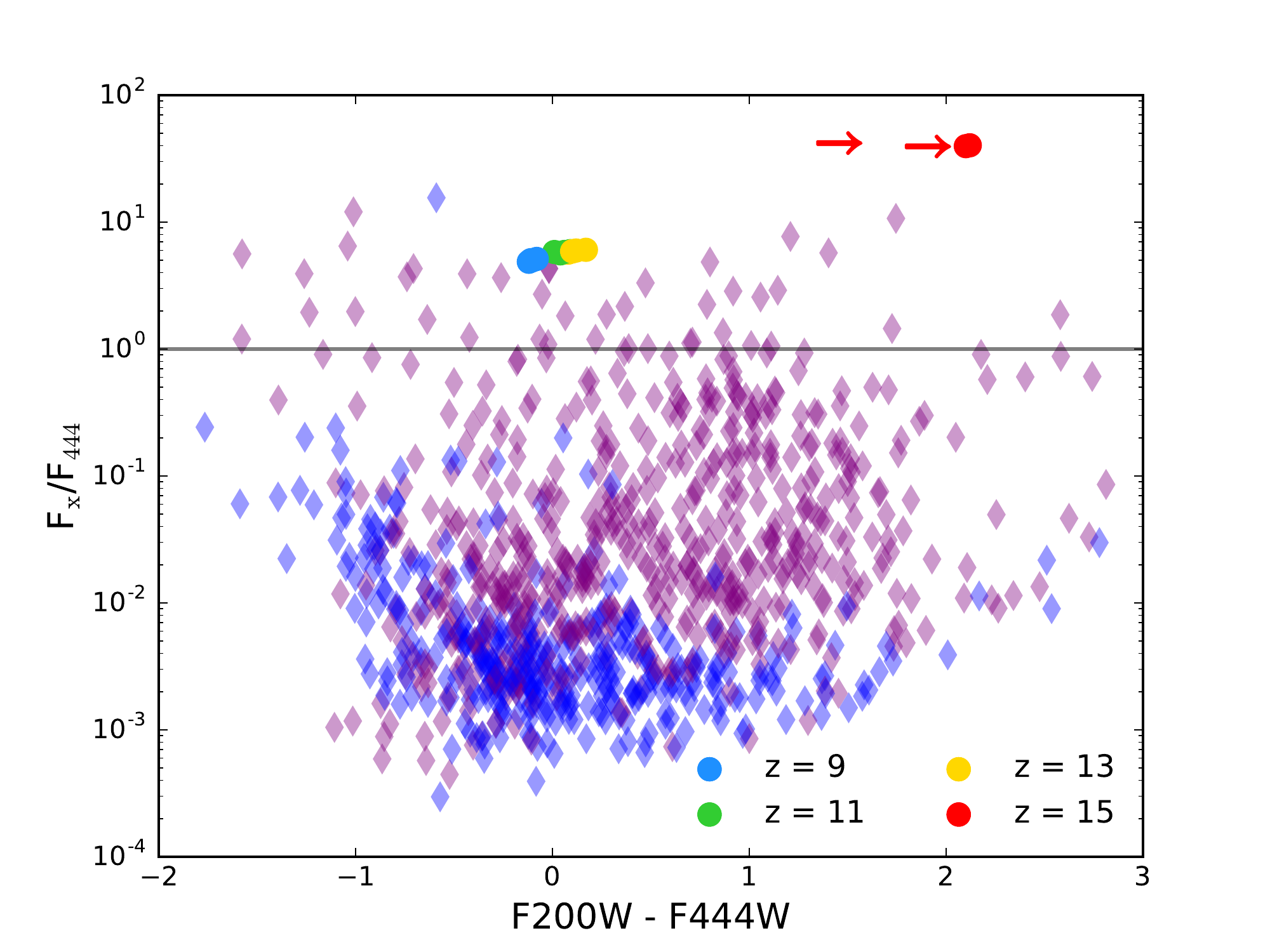}
\caption{Our proposed multi-wavelength selection procedure for identifying OBG candidates accreting via a standard disk with low metallicity, as this is the case with the highest X-ray fluxes: Left Panel - shows the color dropout selection wherein the colored arrows are our template OBGs and the purple arrows are the CANDELS sources. Our model OBG candidates shown as blue, green, red and yellow points are selected at various times all $z \geq 9$. Note that for OBG sources at $z = 15$, the Lyman-break falls within $F_{200W}$ band; Right Panel - the blue diamonds are the sources with simulated X-ray fluxes, the purple ones are CANDELS sources with measured X-ray fluxes. Here we show the X-ray to rest-frame optical flux ratio selection, wherein our OBG template sources, once again shown as blue, green, red and yellow points, can be easily distinguished from star-forming galaxies in CANDELS with and without AGN.}
\label{fig:color-color_selection}
\end{figure}

	We now demonstrate the robustness of our OBG selection criteria using existing deep {\it Hubble}, {\it Spitzer} and {\it Chandra} data. Potential OBG candidates could be lurking in the low luminosity AGN in the CANDELS data-set, and contaminating star-forming galaxies could also be present in the same data. We use the CANDELS-GOODS-S photometric data release version 1.1 \cite{Guo_2013} for the first color-color cut. For each object in the CANDELS catalog, we interpolate the IR photometry into three {\it JWST} filters:  $F_{090W}$, $F_{200W}$, and $F_{444W}$. We pick the $F_{090W}$ photometric band as it allows a redshift selection for all sources at $z > 6$. The two additional bands $F_{200W}$ and $F_{444W}$ are chosen to probe the SED in the NIR and MIR regions, since these sources as expected to have much redder colors than other high-redshift contaminating sources. Given the high expected redshifts for these sources and IGM absorption, these filters are explicitly chosen to detect the Lyman break via the drop-out technique. Then, we determine the X-ray fluxes for all CANDELS sources, both detected and marginally detected faint sources. The marginally detected faint sources are assigned an X-ray flux that is derived using the prescription developed in \cite{nico_2016} and \cite{Basu-Zych_2013}. As per this prescription CANDELS sources that do not have X-ray counter-parts are assigned X-ray fluxes via the relation between SFR, redshift, and X-ray luminosity. Then we apply our multiple selection criteria to this combined catalog; the color-color cut followed by the high $F_{X}/F_{444W}$ cut and over-plot on these diagnostic plots our OBG model template spectra in same filters as shown in Fig.~\ref{fig:color-color_selection}. For the purpose of this exercise, we produce only high-redshift template sources ($z \in \{9,11,13,15\}$) due to the requirement of low metallicity and appropriate Lyman-Werner background, for a range of DCBH seed masses ($M_\bullet/(10^5 \, M_\odot) \in \{0.2, 0.4, 0.6, 0.8, 1.0\}$).

We find that the OBG stage can be isolated with the following criteria, which extend those reported in Pacucci et al. (2016):

\begin{enumerate}
\item{Pre-select sources with bolometric luminosity in excess of $10^{44}\,{\rm erg\,s^{-1}}$}
\item After pre-selection and eliminating blue sources, find dropouts in $F_{090W}$ that are detected in both $F_{200W}$ and $F_{444W}$ with $(F_{090W}  - F_{220W}) > 0$, and $-0.3 < (F_{200W} - F_{444W}) < 0.3$
\item Find an X-ray counterpart and its associated X-ray flux.
\item Isolate sources with a high X-ray to rest-frame optical flux ratio ($F_{X}/F_{444W} >>1$).
\end{enumerate}

In Fig.~\ref{fig:color-color_selection}, we apply the selection criteria to our template sources as well as CANDELS sources with real (measured) and estimated (derived using correlations) X-ray fluxes.  In the left panel, we display a color-color plot of our models along with each of CANDELS sources that would satisfy the first criterion. Depending on the metallicity of the gas and the time at which the spectrum is computed, the IR spectral slope may vary slightly. In this work, we have selected the time slice that produces the highest X-ray flux. In the right panel, we display all sources that would have been detected in CDF-S ($F_x > 6\times 10^{-18} \ \mathrm{erg} \, \mathrm{cm}^{-2} \, \mathrm{s}^{-1}$) {\it but that do not necessarily satisfy our first criterion}.  Using the ratio of X-ray flux to rest-frame optical flux, our OBG sources easily stand out amongst star-forming galaxies and AGN. There are no CANDELS sources in the GOODS-S field that would satisfy all our criteria. We note that the \cite{Pacucci_2016} DCBH candidates are not in our CANDELS catalog as we select sources in the OBG stage during which the X-ray flux is the highest. Our selection of the OBG stage is consistent with the merger driven scenario producing a bright quasar phase. We also explore the dependence of our selection criteria on initial DCBH seed mass and redshift and report that decreasing the DCBH seed mass simply causes the overall flux to decrease, while the X-ray to rest-frame optical flux ratio is independent of seed mass.  There is very little redshift-dependence on the measured colors, due to the fact that these spectra are flat in the IR, until about $z=15$, at which point the Lyman break falls within $F_{200W}$. Sources that satisfy our selection criteria are optimal candidates for spectral follow-up with NIRSPEC and from our modeling work, we predict up to 10 sources within $z = 6 - 10$ in the CANDELS field. For instance, to detect the fiducial OBG continuum at $M_{\rm AB} \sim 26$ (as shown in the top panel of Fig.~6), with NIRSPEC at 3 microns with a $S/N \sim 10$  would take about 3 - 4 hours and the detection of lines would take $\sim$ 1 hr of integration as per the available exposure time calculators
.
\section{Discussion and Future Prospects}

The origin of initial black hole seeds that eventually grow to produce the observed supermassive black holes in the high redshift universe is currently debated. Traditionally it was believed that the end states of the first stars to form - believed to be massive - produced remnants that yielded black hole seeds with masses between $\sim 10 - 100 \msun$. The discovery of several luminous high redshift quasars at $z > 7$ powered by $10^9 \msun$ black holes when the universe was less than 7\% of its present age causes a timing crunch. In order to alleviate this, it was proposed that initial black hole seeds that might form from the direct collapse of un-enriched pre-galactic gas disks might yield seeds with initial masses of $10^{4-5}\,\msun$. These massive DCBH seeds could then grow rapidly and evolve into the bright quasars that are seen. Since black holes erase the initial conditions of their mass assembly, we look to the properties of the host galaxy and environment to test and discriminate between these two models for light and massive seed production in the early universe. In particular, we exploit the unique relationship between the black hole mass and the properties of the host galaxy - a key feature of the DCBH formation process - to discriminate seed formation physics. In terms of the physics, what really distinguishes light BH seeds from massive BH seeds is the epoch at which they have super-boosted growth, if we are to explain the origin of the supermassive black holes that are powering the highest redshift quasars. In order for light seed models to work, brief periods of super-Eddington accretion are required post seed formation; whereas for DCBHs the process of formation itself likely involves super-Eddington accretion. Interestingly, as we show here for DCBHs there is a transitory phase at high redshift, the OBG stage, when there is a unique signature viz-a-viz host galaxy properties that might provide a clue to the initial seed mass. 

Contrary to the local $M_{\rm bh}-M_{\rm bulge}$ relation wherein the central black hole mass is a small fraction, up to $\sim 0.5\%$, of the stellar mass of the bulge, we find that at high redshifts if SMBHs are seeded with massive DCBHs, then they transiently enter an OBG phase, characterized by ${M_{BH}}/{M_*} \sim 1$ \citep{Agarwal_2013,Kormendy_Ho_2013}. In this case, a unique set of detectable observational features are produced. Here we present template multi-wavelength calculations of OBG spectra and show that they are distinguishable from the lower-mass Pop III seeded systems as well as other strong infra-red emitting sources like normal stellar populations at $z \gtrsim 6$ and dusty, old and red galaxies at low redshifts. One note-worthy feature is that the accreting DCBH in an OBG completely outshines the stellar component at X-ray wavelengths $> 0.1\, \mathrm{keV}$, and in the infra-red. This allows us to pre-select viable OBG candidates, as a normal stellar population with a Pop III black hole seed by contrast would be undetected in the X-ray with current data and will also remain undetectable by {\it JWST} due to the significantly lower reprocessed emission in the infra-red. This allows us to immediately sift out infra-red bright OBG candidates for {\it JWST} observations. In the MIRI bands, we find that OBGs can be easily distinguished from contaminants such as nearby quiescent red galaxies, and  high redshift star forming galaxies without a DCBH nucleus. To summarize, we derive three criteria in color-color space $(F_{090W}-F_{200W}) > 0$ and $ -0.3 < (F_{200W}-F_{444W}) < 0.3$ and high ratio of X-ray flux to rest-frame optical flux  ($F_X/F_{444W}>>1$) to select for OBG candidates. 

With the selection criteria derived, the obvious next question to address is the expected abundance of OBGs. Estimates of the expected abundance of both Pop III seeds and DCBH seeds are currently highly uncertain, and this is a topic of active research amongst several groups working on the formation of early black holes that is yet to be fully settled. In a recent paper that utilizes simulations to assess and compute the abundance of the sites including the appropriate conditions - Lyman-Werner flux - required to form DCBH seeds, \cite{Habuzit_2016} estimate that DCBH seeds could account for the origin of all the observed high-redshift quasars at $z >6$. Meanwhile, in a preprint \cite{Pezzulli_2016} note that we might be lacking a proper census in the X-rays for early black holes either due to the uncertainty in the occupation fraction or prevalence of obscured super-Eddington accretion. Therefore, it is patently clear that the IR wavelengths are the most promising as we have shown to obtain demographic properties of this population. The abundance of these sources will of course determine the precise observational strategy to be pursued - either by JWST as we propose here or with shallower, larger area WFIRST surveys. Given the bright IR magnitudes that we estimate here for these template cases, it is clear that WFIRST surveys would be an efficient way forward to pursue to detect OBGs. In future work, we intend to explore these detection strategies in more detail.

	A recent calculation by \cite{Ricotti_2016} reports an expected abundance for Pop III stars of $\sim$ a few hundred per comoving Mpc$^{3}$. These lighter Pop III black hole seeds are expected however to be significantly more abundant than DCBH seeds in the early universe. Using our state-of-the-art theoretical models that cover possible accretion modes and formation times, we predict an abundance with an upper limit range of $\approx 7.5 \times 10^{-4} - 2.5 \times 10^{-5}$ for OBG systems per Mpc$^{3}$. We estimate this abundance by extrapolating the fraction of pristine haloes that are exposed to a critically high LW flux \citep{Dijkstra_2014}. As a first order estimate, using the peak of the critical flux distribution from \cite{Agarwal_2016} to estimate the number of viable hosts for DCBH formation and assembly, in the 800 arcmin$^2$ area of the CANDELS fields which corresponds to a $\sim 9 \times 10^{-3}$ Gpc$^3$ volume, we expect about a hundred potential DCBH formation sites. However, not all these sources will be DCBHs or be in the OBG stage and be detectable by {\it Chandra} since only a fraction of them are expected to be active. Guided by a recent theoretical estimate by \cite{PFVD_2015} that takes into account the duty cycle of DCBHs, we predict up to 10 DCBH sources between $z=6-10$ in the CANDELS fields. This is consistent with the recent study of \cite{Pacucci_2016} in which the authors claimed the possibility of two DCBH candidates in CANDELS/GOODS-S. In future work, we plan to investigate the expected abundance using snapshots from cosmological simulations that include radiative transfer to track DCBH sites in the OBG stage.

As demonstrated in this paper, growing DCBH seeds in the OBG stage would be detectable primarily from their infra-red excess in the {\it JWST} MIRI bands compared to growing Pop III seeds and other contaminating sources. Detection and NIRSPEC follow-up as well as ground-based spectroscopic follow-up using upcoming 30-m class telescopes of these sources during the OBG stage would confirm and improve our understanding of the direct collapse model as a viable channel for seed formation at high redshifts. We also anticipate that WFIRST surveys could feasibly help us detect this population of OBGs. Given the current uncertainties in the Detection of an OBG population will suggest a higher formation efficiency of DCBH seeds and/or a more efficient accretion scenario. Our models suggest that sources that comprise an accreting Pop III seed that has grown to a final black hole mass of $10^7\,\msun$ by $z > 9$ will be to too faint in the infra-red wavelengths to be detectable by {\it JWST}. Therefore, we make the case that OBGs will be booming sources in the MIRI bands and that a sample of detected OBGs would resolve tension between the two SMBH seeding models by offering us a snap-shot of quasars in the early stages of their lives.

\acknowledgments

PN acknowledges support from a Theoretical and Computational Astrophysics Network grant with award number 1332858 from the National Science Foundation. This work was completed at the Aspen Center for Physics, which is supported by National Science Foundation grant PHY-1066293. Useful conversations with Benny Trakhenbrot and Nicholas Stone at the Aspen Black Holes Workshop are gratefully acknowledged. BA and AR acknowledge support from the TCAN grant for a post-doctoral fellowship and a graduate fellowship respectively. This research was supported in part (AF) by the National Science Foundation under Grant No. NSF PHY11-25915. EZ acknowledges funding from the Swedish Research Council (project 2011-5349). NC acknowledges the YCAA Prize Postdoctoral fellowship. NC and FP acknowledge the SAO $Chandra$ grant AR6-17017B and NASA-ADAP grant MA160009.

\bibliographystyle{aasjournal}
\bibliography{ms_final}

\begin{thebibliography}{}
\expandafter\ifx\csname natexlab\endcsname\relax\def\natexlab#1{#1}\fi

\bibitem[{{Abramowicz} {et~al.}(1988){Abramowicz}, {Czerny}, {Lasota}, \&
  {Szuszkiewicz}}]{Abramowicz_1988}
{Abramowicz}, M.~A., {Czerny}, B., {Lasota}, J.~P., \& {Szuszkiewicz}, E. 1988,
  \apj, 332, 646

\bibitem[{{Agarwal} {et~al.}(2014){Agarwal}, {Dalla Vecchia}, {Johnson},
  {Khochfar}, \& {Paardekooper}}]{Agarwal_2014}
{Agarwal}, B., {Dalla Vecchia}, C., {Johnson}, J.~L., {Khochfar}, S., \&
  {Paardekooper}, J.-P. 2014, \mnras, 443, 648

\bibitem[{{Agarwal} {et~al.}(2013){Agarwal}, {Davis}, {Khochfar}, {Natarajan},
  \& {Dunlop}}]{Agarwal_2013}
{Agarwal}, B., {Davis}, A.~J., {Khochfar}, S., {Natarajan}, P., \& {Dunlop},
  J.~S. 2013, \mnras, 432, 3438

\bibitem[{{Agarwal} {et~al.}(2015){Agarwal}, {Johnson}, {Zackrisson}, {Labbe},
  {van den Bosch}, {Natarajan}, \& {Khochfar}}]{Agarwal_2015}
{Agarwal}, B., {Johnson}, J.~L., {Zackrisson}, E., {et~al.} 2015, ArXiv
  e-prints, arXiv:1510.01733

\bibitem[{{Agarwal} {et~al.}(2012){Agarwal}, {Khochfar}, {Johnson}, {Neistein},
  {Dalla Vecchia}, \& {Livio}}]{Agarwal_2012}
{Agarwal}, B., {Khochfar}, S., {Johnson}, J.~L., {et~al.} 2012, \mnras, 425,
  2854

\bibitem[{{Agarwal} {et~al.}(2016){Agarwal}, {Smith}, {Glover}, {Natarajan}, \&
  {Khochfar}}]{Agarwal_2016}
{Agarwal}, B., {Smith}, B., {Glover}, S., {Natarajan}, P., \& {Khochfar}, S.
  2016, \mnras, arXiv:1504.04042

\bibitem[{{Alexander} \& {Natarajan}(2014)}]{TalPN_2014}
{Alexander}, T., \& {Natarajan}, P. 2014, Science, 345, 1330

\bibitem[{{Aykutalp} {et~al.}(2014){Aykutalp}, {Wise}, {Spaans}, \&
  {Meijerink}}]{Aykutalp_2014}
{Aykutalp}, A., {Wise}, J.~H., {Spaans}, M., \& {Meijerink}, R. 2014, \apj,
  797, 139

\bibitem[{{Basu-Zych} {et~al.}(2013){Basu-Zych}, {Lehmer}, {Hornschemeier},
  {Bouwens}, {Fragos}, {Oesch}, {Belczynski}, {Brandt}, {Kalogera}, {Luo},
  {Miller}, {Mullaney}, {Tzanavaris}, {Xue}, \& {Zezas}}]{Basu-Zych_2013}
{Basu-Zych}, A.~R., {Lehmer}, B.~D., {Hornschemeier}, A.~E., {et~al.} 2013,
  \apj, 762, 45

\bibitem[{{Begelman} {et~al.}(2006){Begelman}, {Volonteri}, \&
  {Rees}}]{Begelman_2006}
{Begelman}, M.~C., {Volonteri}, M., \& {Rees}, M.~J. 2006, \mnras, 370, 289

\bibitem[{{Bromm} \& {Loeb}(2003)}]{Bromm_Loeb_2003}
{Bromm}, V., \& {Loeb}, A. 2003, \apj, 596, 34

\bibitem[{{Cappelluti} {et~al.}(2016){Cappelluti}, {Comastri}, {Fontana},
  {Zamorani}, {Amorin}, {Castellano}, {Merlin}, {Santini}, {Elbaz},
  {Schreiber}, {Shu}, {Wang}, {Dunlop}, {Bourne}, {Bruce}, {Buitrago},
  {Micha{\l}owski}, {Derriere}, {Ferguson}, {Faber}, \& {Vito}}]{nico_2016}
{Cappelluti}, N., {Comastri}, A., {Fontana}, A., {et~al.} 2016, \apj, 823, 95

\bibitem[{{Choi} {et~al.}(2013){Choi}, {Shlosman}, \& {Begelman}}]{Choi_2013}
{Choi}, J.-H., {Shlosman}, I., \& {Begelman}, M.~C. 2013, \apj, 774, 149

\bibitem[{{Decarli} {et~al.}(2012){Decarli}, {Walter}, {Yang}, {Carilli},
  {Fan}, {Hennawi}, {Kurk}, {Riechers}, {Rix}, {Strauss}, \&
  {Venemans}}]{Decarli_2012}
{Decarli}, R., {Walter}, F., {Yang}, Y., {et~al.} 2012, \apj, 756, 150

\bibitem[{{Devecchi} \& {Volonteri}(2009)}]{devecchi_2009}
{Devecchi}, B., \& {Volonteri}, M. 2009, \apj, 694, 302

\bibitem[{{Dijkstra} {et~al.}(2014){Dijkstra}, {Ferrara}, \&
  {Mesinger}}]{Dijkstra_2014}
{Dijkstra}, M., {Ferrara}, A., \& {Mesinger}, A. 2014, \mnras, 442, 2036

\bibitem[{{Eisenstein} \& {Loeb}(1995)}]{Eisenstein_Loeb_1995}
{Eisenstein}, D.~J., \& {Loeb}, A. 1995, \apj, 443, 11

\bibitem[{{Fan} {et~al.}(2001)}]{Fan_2001}
{Fan}, X., {et~al.} 2001, \aj, 122, 2833

\bibitem[{{Ferland} {et~al.}(2013){Ferland}, {Porter}, {van Hoof}, {Williams},
  {Abel}, {Lykins}, {Shaw}, {Henney}, \& {Stancil}}]{Cloudy}
{Ferland}, G.~J., {Porter}, R.~L., {van Hoof}, P.~A.~M., {et~al.} 2013,
  {Revista Mexicana de Astronomia y Astrofisica}, 49, 137

\bibitem[{{Ferrara} {et~al.}(2014){Ferrara}, {Salvadori}, {Yue}, \&
  {Schleicher}}]{Ferrara_2014}
{Ferrara}, A., {Salvadori}, S., {Yue}, B., \& {Schleicher}, D. 2014, \mnras,
  443, 2410

\bibitem[{{Ferrarese} \& {Merritt}(2000)}]{Ferrarese_Merritt_2000}
{Ferrarese}, L., \& {Merritt}, D. 2000, \apjl, 539, L9

\bibitem[{{Guo} {et~al.}(2013){Guo}, {Ferguson}, {Giavalisco}, {Barro},
  {Willner}, {Ashby}, {Dahlen}, {Donley}, {Faber}, {Fontana}, {Galametz},
  {Grazian}, {Huang}, {Kocevski}, {Koekemoer}, {Koo}, {McGrath}, {Peth},
  {Salvato}, {Wuyts}, {Castellano}, {Cooray}, {Dickinson}, {Dunlop}, {Fazio},
  {Gardner}, {Gawiser}, {Grogin}, {Hathi}, {Hsu}, {Lee}, {Lucas}, {Mobasher},
  {Nandra}, {Newman}, \& {van der Wel}}]{Guo_2013}
{Guo}, Y., {Ferguson}, H.~C., {Giavalisco}, M., {et~al.} 2013, \apjs, 207, 24

\bibitem[{{Habouzit} {et~al.}(2016){Habouzit}, {Volonteri}, {Latif}, {Dubois},
  \& {Peirani}}]{Habuzit_2016}
{Habouzit}, M., {Volonteri}, M., {Latif}, M., {Dubois}, Y., \& {Peirani}, S.
  2016, \mnras, 463, 529

\bibitem[{{H{\"a}ring} \& {Rix}(2004)}]{Haring_2004}
{H{\"a}ring}, N., \& {Rix}, H.-W. 2004, \apjl, 604, L89

\bibitem[{{Johnson} {et~al.}(2012){Johnson}, {Whalen}, {Fryer}, \&
  {Li}}]{Johnson_2012}
{Johnson}, J.~L., {Whalen}, D.~J., {Fryer}, C.~L., \& {Li}, H. 2012, \apj, 750,
  66

\bibitem[{{Kormendy} \& {Ho}(2013)}]{Kormendy_Ho_2013}
{Kormendy}, J., \& {Ho}, L.~C. 2013, \araa, 51, 511

\bibitem[{{Koushiappas} {et~al.}(2004){Koushiappas}, {Bullock}, \&
  {Dekel}}]{Koushiappas_2004}
{Koushiappas}, S.~M., {Bullock}, J.~S., \& {Dekel}, A. 2004, \mnras, 354, 292

\bibitem[{{Kroupa}(2001)}]{Kroupa_2001}
{Kroupa}, P. 2001, \mnras, 322, 231

\bibitem[{{Latif} \& {Ferrara}(2016)}]{Latif_2016}
{Latif}, M.~A., \& {Ferrara}, A. 2016, ArXiv e-prints, arXiv:1605.07391

\bibitem[{{Latif} {et~al.}(2013){Latif}, {Schleicher}, {Schmidt}, \&
  {Niemeyer}}]{Latif_2013}
{Latif}, M.~A., {Schleicher}, D.~R.~G., {Schmidt}, W., \& {Niemeyer}, J. 2013,
  \mnras, 433, 1607

\bibitem[{{Leitherer} {et~al.}(1999){Leitherer}, {Schaerer}, {Goldader},
  {Delgado}, {Robert}, {Kune}, {de Mello}, {Devost}, \&
  {Heckman}}]{Leitherer_1999}
{Leitherer}, C., {Schaerer}, D., {Goldader}, J.~D., {et~al.} 1999, \apjs, 123,
  3

\bibitem[{{Lodato} \& {Natarajan}(2006)}]{Lodato_Natarajan_2006}
{Lodato}, G., \& {Natarajan}, P. 2006, \mnras, 371, 1813

\bibitem[{{Lodato} \& {Natarajan}(2007)}]{Lodato_Natarajan_2007}
---. 2007, \mnras, 377, L64

\bibitem[{{Machacek} {et~al.}(2001){Machacek}, {Bryan}, \&
  {Abel}}]{Machacek_2001}
{Machacek}, M.~E., {Bryan}, G.~L., \& {Abel}, T. 2001, \apj, 548, 509

\bibitem[{{McKinney} {et~al.}(2014){McKinney}, {Tchekhovskoy}, {Sadowski}, \&
  {Narayan}}]{McKinney_2014}
{McKinney}, J.~C., {Tchekhovskoy}, A., {Sadowski}, A., \& {Narayan}, R. 2014,
  \mnras, 441, 3177

\bibitem[{{Mortlock} {et~al.}(2011){Mortlock}, {Warren}, {Venemans}, {Patel},
  {Hewett}, {McMahon}, {Simpson}, {Theuns}, {Gonz{\'a}les-Solares}, {Adamson},
  {Dye}, {Hambly}, {Hirst}, {Irwin}, {Kuiper}, {Lawrence}, \&
  {R{\"o}ttgering}}]{Mortlock_2011}
{Mortlock}, D.~J., {Warren}, S.~J., {Venemans}, B.~P., {et~al.} 2011, \nat,
  474, 616

\bibitem[{{Natarajan}(2014)}]{PN_2014}
{Natarajan}, P. 2014, General Relativity and Gravitation, 46, 1702

\bibitem[{{Natarajan} \& {Volonteri}(2012)}]{Natarajan_Volonteri_2012}
{Natarajan}, P., \& {Volonteri}, M. 2012, \mnras, 422, 2051

\bibitem[{{Oesch} {et~al.}(2016){Oesch}, {Brammer}, {van Dokkum},
  {Illingworth}, {Bouwens}, {Labbe}, {Franx}, {Momcheva}, {Ashby}, {Fazio},
  {Gonzalez}, {Holden}, {Magee}, {Skelton}, {Smit}, {Spitler}, {Trenti}, \&
  {Willner}}]{Oesch_2016}
{Oesch}, P.~A., {Brammer}, G., {van Dokkum}, P.~G., {et~al.} 2016, ArXiv
  e-prints, arXiv:1603.00461

\bibitem[{{Oh} \& {Haiman}(2002)}]{Oh_2002}
{Oh}, S.~P., \& {Haiman}, Z. 2002, \apj, 569, 558

\bibitem[{{Omukai}(2001)}]{Omukai_2001}
{Omukai}, K. 2001, \apj, 546, 635

\bibitem[{{O'Shea} \& {Norman}(2008)}]{Oshea_2008}
{O'Shea}, B.~W., \& {Norman}, M.~L. 2008, \apj, 673, 14

\bibitem[{{Pacucci} \& {Ferrara}(2015)}]{Pacucci_2015}
{Pacucci}, F., \& {Ferrara}, A. 2015, \mnras, 448, 104

\bibitem[{{Pacucci} {et~al.}(2016){Pacucci}, {Ferrara}, {Grazian}, {Fiore},
  {Giallongo}, \& {Puccetti}}]{Pacucci_2016}
{Pacucci}, F., {Ferrara}, A., {Grazian}, A., {et~al.} 2016, \mnras, 459, 1432

\bibitem[{{Pacucci} {et~al.}(2015{\natexlab{a}}){Pacucci}, {Ferrara},
  {Volonteri}, \& {Dubus}}]{PFVD_2015}
{Pacucci}, F., {Ferrara}, A., {Volonteri}, M., \& {Dubus}, G.
  2015{\natexlab{a}}, \mnras, 454, 3771

\bibitem[{{Pacucci} {et~al.}(2015{\natexlab{b}}){Pacucci}, {Volonteri}, \&
  {Ferrara}}]{PVF_2015}
{Pacucci}, F., {Volonteri}, M., \& {Ferrara}, A. 2015{\natexlab{b}}, \mnras,
  452, 1922

\bibitem[{{Paczynski} \& {Abramowicz}(1982)}]{Paczynski_1982}
{Paczynski}, B., \& {Abramowicz}, M.~A. 1982, \apj, 253, 897

\bibitem[{{Park} {et~al.}(2016){Park}, {Ricotti}, {Natarajan},
  {Bogdanovi{\'c}}, \& {Wise}}]{Park_2016}
{Park}, K., {Ricotti}, M., {Natarajan}, P., {Bogdanovi{\'c}}, T., \& {Wise},
  J.~H. 2016, \apj, 818, 184

\bibitem[{{Pezzulli} {et~al.}(2016){Pezzulli}, {Valiante}, {Orofino},
  {Schneider}, {Gallerani}, \& {Sbarrato}}]{Pezzulli_2016}
{Pezzulli}, E., {Valiante}, R., {Orofino}, M.~C., {et~al.} 2016, ArXiv
  e-prints, arXiv:1612.04188

\bibitem[{{Raiter} {et~al.}(2010){Raiter}, {Schaerer}, \&
  {Fosbury}}]{Raiter_2010}
{Raiter}, A., {Schaerer}, D., \& {Fosbury}, R.~A.~E. 2010, \aap, 523, A64

\bibitem[{{Regan} \& {Haehnelt}(2009)}]{Regan_2009}
{Regan}, J.~A., \& {Haehnelt}, M.~G. 2009, \mnras, 396, 343

\bibitem[{{Ricotti}(2016)}]{Ricotti_2016}
{Ricotti}, M. 2016, \mnras, 462, 601

\bibitem[{{Sadowski}(2009)}]{Sadowski_2009}
{Sadowski}, A. 2009, \apjs, 183, 171

\bibitem[{{Shakura} \& {Sunyaev}(1976)}]{Shakura_Sunyaev_1976}
{Shakura}, N.~I., \& {Sunyaev}, R.~A. 1976, \mnras, 175, 613

\bibitem[{{Shang} {et~al.}(2010){Shang}, {Bryan}, \& {Haiman}}]{Shang_2010}
{Shang}, C., {Bryan}, G.~L., \& {Haiman}, Z. 2010, \mnras, 402, 1249

\bibitem[{{Tanaka} {et~al.}(2012){Tanaka}, {Perna}, \& {Haiman}}]{Tanaka_2012}
{Tanaka}, T., {Perna}, R., \& {Haiman}, Z. 2012, \mnras, 425, 2974

\bibitem[{{Toomre}(1964)}]{Toomre_1964}
{Toomre}, A. 1964, \apj, 139, 1217

\bibitem[{{Treister} {et~al.}(2013){Treister}, {Schawinski}, {Volonteri}, \&
  {Natarajan}}]{ET_2013}
{Treister}, E., {Schawinski}, K., {Volonteri}, M., \& {Natarajan}, P. 2013,
  \apj, 778, 130

\bibitem[{{Tremaine} {et~al.}(2002){Tremaine}, {Gebhardt}, {Bender}, {Bower},
  {Dressler}, {Faber}, {Filippenko}, {Green}, {Grillmair}, {Ho}, {Kormendy},
  {Lauer}, {Magorrian}, {Pinkney}, \& {Richstone}}]{Tremaine_2002}
{Tremaine}, S., {Gebhardt}, K., {Bender}, R., {et~al.} 2002, \apj, 574, 740

\bibitem[{{Volonteri}(2012)}]{MV_2012}
{Volonteri}, M. 2012, Science, 337, 544

\bibitem[{{Volonteri} {et~al.}(2008){Volonteri}, {Lodato}, \&
  {Natarajan}}]{Volonteri_2008}
{Volonteri}, M., {Lodato}, G., \& {Natarajan}, P. 2008, \mnras, 383, 1079

\bibitem[{{Weigel} {et~al.}(2015){Weigel}, {Schawinski}, {Treister}, {Urry},
  {Koss}, \& {Trakhtenbrot}}]{Weigel_2015}
{Weigel}, A.~K., {Schawinski}, K., {Treister}, E., {et~al.} 2015, \mnras, 448,
  3167

\bibitem[{{Wu} {et~al.}(2015){Wu}, {Wang}, {Fan}, {Yi}, {Zuo}, {Bian}, {Jiang},
  {McGreer}, {Wang}, {Yang}, {Yang}, {Thompson}, \& {Beletsky}}]{Wu_2015}
{Wu}, X.-B., {Wang}, F., {Fan}, X., {et~al.} 2015, \nat, 518, 512

\bibitem[{{Zackrisson} {et~al.}(2011){Zackrisson}, {Rydberg}, {Schaerer},
  {{\"O}stlin}, \& {Tuli}}]{Zackrisson_2011}
{Zackrisson}, E., {Rydberg}, C.-E., {Schaerer}, D., {{\"O}stlin}, G., \&
  {Tuli}, M. 2011, \apj, 740, 13

\end{thebibliography}
\end{document}